\documentclass[11pt,a4paper]{article}
\usepackage{jcappub}
\usepackage[font=footnotesize,up]{caption}
\usepackage{subcaption}
\usepackage{wrapfig}
\usepackage{enumerate}
\usepackage{multirow}
\usepackage{booktabs}
\usepackage{array}
\usepackage{tabulary}
\usepackage[normalem]{ulem}

\title{Optimized analysis method for indirect dark matter searches with Imaging Air Cherenkov Telescopes}

\author[a]{J. Aleksi\'c,}
\author[b,a]{J. Rico}
\author[a]{and M. Martinez}

\affiliation[a]{Institut de F\'isica d'Altes Energies (IFAE),\\
Campus UAB, E-08193 Bellaterra, Spain}
\affiliation[b]{Instituci\'o Catalana de Recerca i Estudis Avan\c{c}ats (ICREA), \\
E-08010 Barcelona, Spain}

\emailAdd{jelena@ifae.es}
\emailAdd{jrico@ifae.es}
\emailAdd{martinez@ifae.es}

\abstract{We propose a dedicated analysis approach for indirect Dark Matter searches with Imaging Air 
Cherenkov Telescopes. By using the full likelihood analysis, we take complete advantage of the distinct 
features expected in the gamma ray spectrum of Dark Matter origin, achieving better sensitivity with 
respect to the standard analysis chains. We describe the method and characterize its general performance. 
We also compare its sensitivity with that of the current standards for several Dark Matter annihilation 
models, obtaining gains of up to factors of order of 10. We compute the improved limits that can be 
reached using this new approach, taking as an example existing estimates for several benchmark models 
as well as the recent results from VERITAS on observations of the dwarf spheroidal galaxy Segue 1. 
Furthermore, we estimate the sensitivity of Cherenkov Telescopes for monochromatic line signals. 
Predictions are made on improvement that can be achieved for MAGIC and CTA. Lastly, we discuss how 
this method can be applied in a global, sensitivity-optimized indirect Dark Matter search that combines 
the results of all Cherenkov observatories of the present generation.}


\begin{document}
\maketitle
\flushbottom

\section{Introduction}
\label{Sec1}

The existence of Dark Matter (DM) has been confirmed by observational evidence on all scales, yet its nature still remains a 
mystery (for a review, see e.g. \cite{1.1}). Among theories that try to describe DM and incorporate its presence in our image 
of the Universe today, the Cold Dark Matter paradigm offers the most satisfactory explanation. It requires the DM particle 
to be cold, neutral, stable on cosmological scales, consistent with the Big Bang nucleosynthesis and not excluded by the 
existing experimental constraints \cite{1.2}.

Among the possible candidates complying with these conditions, the Weakly Interacting Massive Particles (WIMPs) are the 
most widely considered. However, such particles do not exist within the framework of the Standard Model (SM), so one 
must go beyond its limits to look for WIMPs. For example, the Supersymmetric extension of the SM \cite{1.3} suggests 
various natural DM particle candidates, with lightest neutralino being the most studied one. Assuming that neutralinos can 
self-annihilate, the resulting by-products are expected to be SM particles detectable from Earth, like electrons, positrons, 
photons and neutrinos.

According to various models (see, e.g., \cite{1.4}), the DM particle mass is expected to be in the few GeV - few TeV energy 
range; therefore, highly energetic photons resulting from WIMP annihilation might be visible to the Imaging Air Cherenkov 
Telescopes (IACTs). Gamma rays are especially attractive from the point of view of indirect DM searches: not only do they 
trace back to the place of their creation and can be detected from space and ground, their energy pattern also preserves 
information on the DM particle they originated from. Spectral features like cut-off \cite{1.5}, monochromatic line \cite{1.6} 
or spectral hardening due to the internal bremsstrahlung \cite{1.7} cannot be imitated by the conventional astrophysical 
sources; as such, they are considered to be the 'smoking guns' of DM detection.

Cherenkov Telescopes have limited duty cycles and great variety of scientific objectives competing for the observation 
time. Their physics programs are primarily focused on detection and study of astrophysical sources, with fundamental 
physics and cosmological issues frequently left on the sidelines. As a consequence, standard analysis tools and methods 
used to process the observed data are usually optimized for sources with, in the majority of cases, featureless spectral 
distributions well described by a simple power law.

Such analysis is suboptimal for DM searches, therefore, we propose an improved, dedicated approach of optimized 
sensitivity for spectral features of DM origin. In comparison with some previous works, that addressed the issue by 
focusing only on the regions of the spectrum with the most peculiar, DM-induced features (see, e.g. \cite {1.8} and 
references within), the method we are suggesting takes full profit of all spectral differences between the signal and the 
background.

We describe the proposed approach in detail in section \ref{Sec2} and then go on to characterize it and compare its 
performance with respect to that of the standard IACT analysis in section \ref{Sec3}. In section \ref{Sec4}, we apply the new 
method to several models of DM emission, show the improvement achievable with respect to the recently published 
experimental searches and make observability predictions for different annihilation channels and instruments. Lastly, 
section \ref{Sec5} is reserved for the discussion and concluding remarks.

\section{Full likelihood method}
\label{Sec2}

In the standard analysis chain of IACTs, the existence of a source is established by a mere comparison of the integrated 
number of events detected in the source region (\textit{n}) with the number of events detected from the control, 
background region(s) (\textit{m}). Both \textit{n} and \textit{m} are random variables that obey Poisson statistics; therefore, 
the number of gamma-ray (\textit{g}) and background (\textit{b}) events present in the source region can be estimated by 
maximization of the following likelihood function \cite{2.1}:
\begin{equation} 
\label{eq1}
\mathcal{L}(g, b | n, m) =
\frac{(g+b)^{n}}{n!}e^{-(g+b)}\times\frac{(\tau b)^m}{m!}e^{-\tau b},
\end{equation} 
where $\tau$ is the normalisation between the signal and background regions (e.g. ratio of their associated observation 
times). This, Poisson likelihood to which we also refer as the \textit{``conventional''} likelihood approach, is what is 
currently used in the standard analysis of the IACT data. Whilst acceptable for sources of astrophysical origin, this method 
does not make any distinction of the potential features present in the gamma-ray spectrum, and as such, it is suboptimal 
for the DM searches.

We propose the use of an alternative, more DM-oriented approach: by making an a priori assumption on the expected 
spectral shape (which is fixed and known for a given DM model), and including it in the maximum likelihood analysis, we 
can completely exploit the spectral information of the events from DM annihilation and achieve better sensitivity with 
respect to the conventional method. This \textit{full likelihood} function has, for a given DM model $M$ with parameters 
$\boldsymbol\theta$, the following form:
\begin{equation} 
\label{eq2}
\mathcal{L}(N_{EST}, M(\boldsymbol\theta)|N_{OBS}, E_1, ..., E_{N_{OBS}})
= \frac{{N_{EST}}^{N_{OBS}}}{N_{OBS}!}e^{-N_{EST}} \times
\prod\limits_{i=1}^{N_{OBS}}\mathcal{P}(E_i; M(\boldsymbol\theta)),
\end{equation} 
with $N_{OBS} ( = n + m)$ and $N_{EST}$ denoting the total number of observed and estimated events, respectively, in  source 
and background regions.

$\mathcal{P}(E_{i}; M(\boldsymbol\theta))$ is the value of the probability density function (PDF) of the event $i$ with 
\textit{measured} energy $E_{i}$. In general, $\mathcal{P}$ may also depend on the measured arrival time and direction of 
the photon, which would reflect on the sensitivity to gamma rays with distinct temporal and spatial structures, 
respectively. However, signals from DM annihilation are expected to be steady, allowing us to integrate out the time in our 
analysis. As for the spatial signatures, they may have a more important role for, e.g., galaxy clusters \cite{2.2}, as they can 
be predicted from halo simulations, although, usually with great uncertainties (see, for example, \cite{2.3, 2.4, 2.5}). 
However, in this work we concentrate mainly on source-candidates that are of angular size smaller or comparable to the 
typical angular resolution of the IACTs ($\sim0.1^{\circ}$) - hence,  we do not expect a contribution from a likelihood 
function dependent on the direction, and we integrate it out as well. Therefore, we define the PDF as a function of 
measured energy only:
\begin{equation}
\mathcal{P}(E; M(\boldsymbol\theta)) = \frac{P (E;
M(\boldsymbol\theta))}{\int\limits_{E_{min}}^{E_{max}}P (E;
M(\boldsymbol\theta)) dE},
\label{eq3}
\end{equation}
where $E_{min}$ and $E_{max}$ are the lower and upper limits of the considered energy range;  $P(E; M(\boldsymbol\theta))$ 
represents the differential rate of signal and background events, such that:
\begin{equation}
P(E; M(\boldsymbol\theta)) =
\begin{cases} P_{B}(E_i), & i \in B \\ P_{S}(E_i;
M(\boldsymbol\theta)), & i \in S
  \end{cases},
\label{eq4}
\end{equation}
with $P_B(E)$ and $P_S(E; M(\boldsymbol\theta))$ being the expected differential rates from the background ($B$) and 
source ($S$) regions, respectively:
\begin{equation}
P_B(E) = \tau \int\limits_0^\infty \frac{d\Phi_B}{dE'}R_B(E; E')dE'
\label{eq5}
\end{equation}
and
\begin{equation}
P_S(E; M(\boldsymbol\theta)) = \int\limits_0^\infty
\frac{d\Phi_B}{dE'}R_B(E; E')dE' + \int\limits_0^\infty
\frac{d\Phi_G(M(\boldsymbol\theta))}{dE'}R_G(E; E')dE'.
\label{eq6}
\end{equation}
True energy is denoted with $E'$; $d\Phi_B$/$dE'$ and $d\Phi_G$/$dE'$ are the differential fluxes of cosmic (background) 
and gamma-ray (signal) radiations, and $R_B(E; E')$ and $R_G(E; E')$ are the telescope response functions to each of them. 
$d\Phi_G$/$dE'$ contains the dependencies on the model parameters $(\boldsymbol\theta)$. 
\begin{figure}[t] \vspace{-20pt} \centering
  \includegraphics[width=0.89\textwidth]{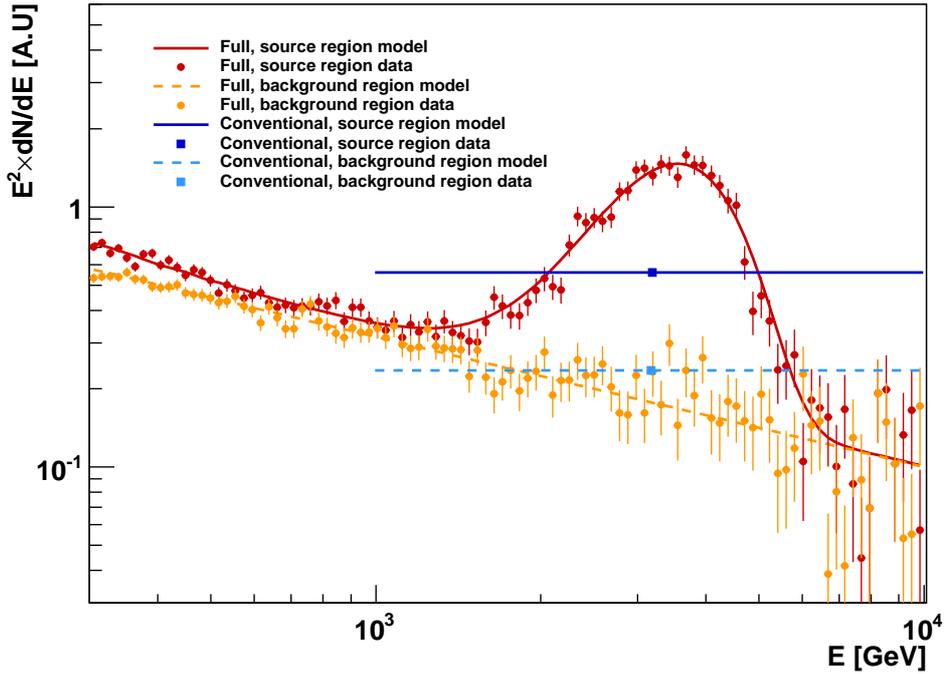}
  \vspace{-5pt}
  \caption{Illustration of the advantage of the full likelihood method 
    with respect to the conventional one. Red and orange lines show 
    the assumed spectral energy distributions of the source and 
    background regions, respectively, while the data points, with the 
    same color code, represent the measured events (fine binning is 
    used for demonstration purposes only, the full likelihood is 
    unbinned). The levels of horizontal blue and cyan lines correspond 
    to the average value within the energy range considered in the 
    conventional method, with dots referring to the measurements. 
    See the main text for more details.}
  \label{Fig1}
  \vspace{-5pt}
\end{figure}

However, in practice, $R_{B}$ can be different for source and background regions, due to its dependence on the  direction of 
the incoming particles within the observed field of view. Such discrepancies are measurable by the telescopes with 
relatively high precision, and the residual statistical and systematic uncertainties can be taken into account in the 
likelihood function through inclusion of the relevant nuisance parameters (see, e.g. \cite{2.1,2.6}). In this work we consider 
that $R_{B}$ is equal in eq.(\ref{eq5}) and eq.(\ref{eq6}) and known with perfect precision. Nevertheless, in section 
\ref{Sec3.4} we evaluate the impact its uncertainties may have on our results.

Apart from the shape of the signal spectral distribution, the given model $M(\boldsymbol\theta)$ also predicts the 
expected number of detected events for a given observation time $T_{OBS}$:
\begin{equation}
N_{EST} = T_{OBS} \int\limits_{E_{min}}^{E_{max}}P(E; M(\boldsymbol\theta))dE,
\label{eq7}
\end{equation}
included in the full likelihood (eq.(\ref{eq2})) through the Poissonian term. 

Lastly, for the comparison of the full with conventional analysis, it is worthwhile to note that 
\begin{equation}
b = \frac{T_{OBS}}{\tau}\int\limits_{E_{min}}^{E_{max}}P_B(E)dE
\label{eq8}
\end{equation}
and
\begin{equation}
g(\boldsymbol\theta) = T_{OBS} \int\limits_{E_{min}}^{E_{max}}P_S(E; M(\boldsymbol\theta))dE - b.
\label{eq9}
\end{equation}

Figure \ref{Fig1} illustrates the advantage of the full likelihood with respect to the conventional one. Both methods are 
based on comparisons of the collected data with the predictions from the signal and background models. The conventional 
method integrates the spectral information in a pre-optimized energy range (for details, see section \ref{Sec3.2}), so that 
the only information used is that of the expected and measured \emph{number of events}. On the other hand, the full 
likelihood compares the expected and measured \emph{energy distributions}, thus fully profiting from the potential 
presence of DM spectral features.

\section{Characterization of the full likelihood method}
\label{Sec3}

In this section, in order to evaluate the performance of the full likelihood concept in the IACT analysis, we test it using fast 
simulations produced under a predefined set of conditions, and compare the results with those of the conventional method 
obtained under the exact same circumstances.

\subsection{The setup}
\label{Sec3.1}

\paragraph{Response Function.}The response functions $R_B$ and $R_G$ of an IACT are governed by its hardware design, 
reconstruction algorithms, selection criteria for quality of the events and for discrimination between the signal and 
background. They are computed by means of measurements and full Monte Carlo simulations, and typically can be 
represented as a product of three factors: effective area $A_{eff}(E', \hat{p}', t)$, angular ($\Sigma(\hat{p}; E', \hat{p}', t)$) and 
energy ($G(E; E', \hat{p}', t)$) reconstruction functions, with $\hat{p}$ and $\hat{p}'$ referring to the measured and true 
directions of the incoming particle, and $t$ to the time of the detection. As mentioned before, the spatial and temporal 
dependencies are integrated out in our analysis, so the response functions we use in this work have the form:
\begin{equation}
R_{B, G}(E; E') = A_{{eff}_{B, G}}(E') G_{B, G}(E; E').
\label{eq10}
\end{equation}

The effective area $A_{eff}$ is the area in which air showers can be observed by the instrument folded with the efficiency of 
all the cuts applied in the analysis. Cherenkov telescopes are not equally sensitive to gamma and cosmic ray showers, and 
effective areas for different particles are different as well.

The energy resolution $\sigma$ is defined as the width of a Gaussian fit to the $(E-E')/E'$ distribution, while the mean of 
that fit is the relative energy bias $\mu$. However, in the real data analysis, the energy reconstruction function might need 
to be described by a more accurate parametrization. 

For the characterization of the full likelihood method in the following tests, as representative response function of a 
current-generation IACT, we use the corresponding functions of the MAGIC 
Telescopes\footnote{http://magic.mppmu.mpg.de} \cite{3.1.1}.

\paragraph{Spectral Functions.}The background emission is produced by the cosmic rays, with a flux well described by a 
simple power law: 
\begin{equation}
\frac{d\Phi_B}{dE'} = A_{B}{E'}^{-\alpha}, 
\label{eq11}
\end{equation}
with spectral index $\alpha$ and intensity $A_{B}$. In practice, however, we only need the value of $P_{B}(E)$ (eq.(\ref{eq5})), 
which is directly measured by the IACTs (or computed from Monte Carlo simulations for projected instruments).

Regarding the spectral form of the signal emission, at this stage, we consider two simple cases:
\begin{enumerate}[(a)]
\item power law ($PL$) of spectral index $\gamma$ and intensity $A_{PL}$: 
\begin{equation}
\frac{d\Phi_G}{dE'} = A_{PL}E'^{-\gamma};
\label{eq12}
\end{equation}
\item a monochromatic line ($L$) at energy $l$ and of intensity $A_{L}$: 
\begin{equation}
\frac{d\Phi_G}{dE'} = A_{L}\delta(E' - l).
\label{eq13}
\end{equation}
\end{enumerate}

\begin{figure} \vspace{-20pt} \centering
  \includegraphics[width=0.49\textwidth]{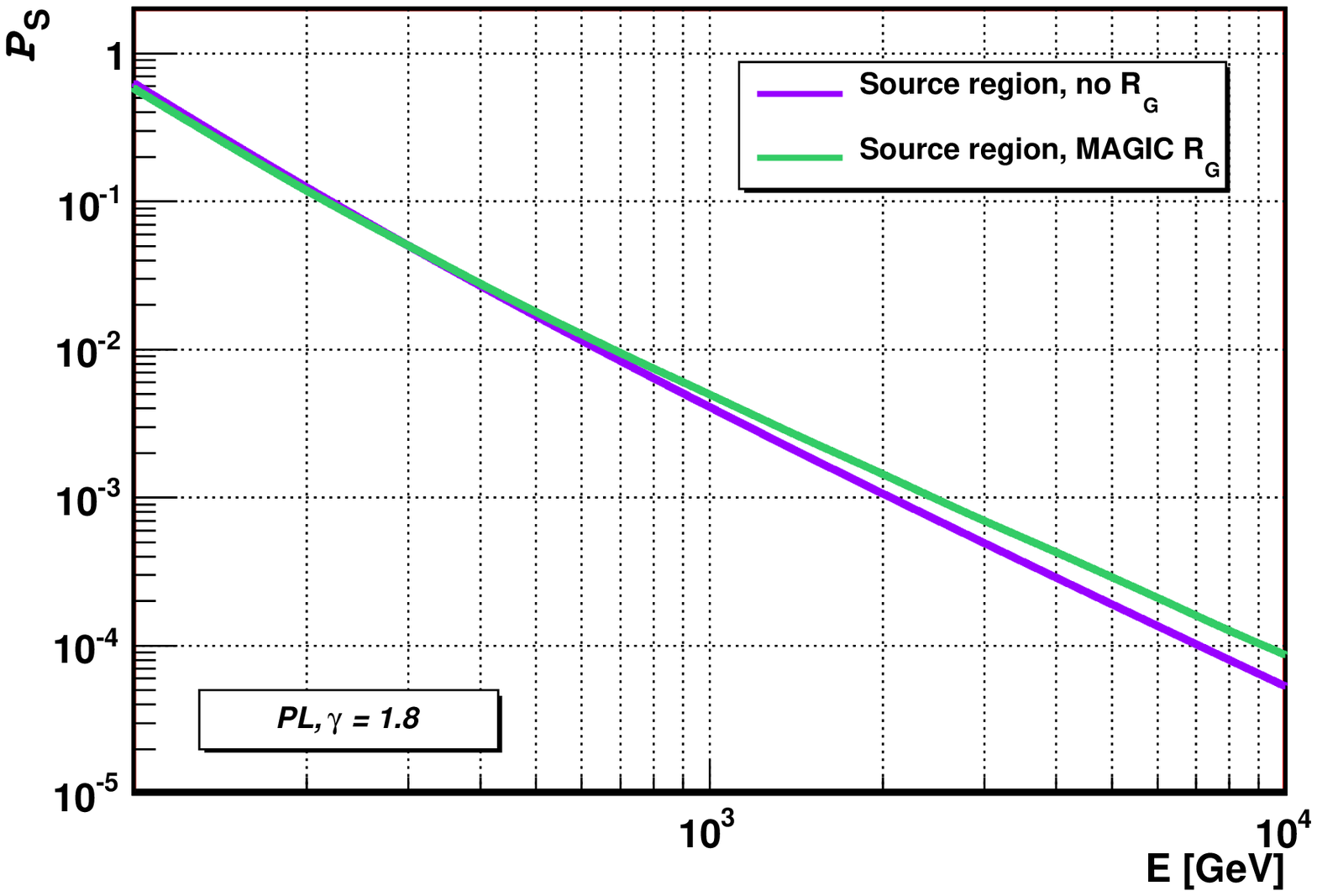}
  \includegraphics[width=0.49\textwidth]{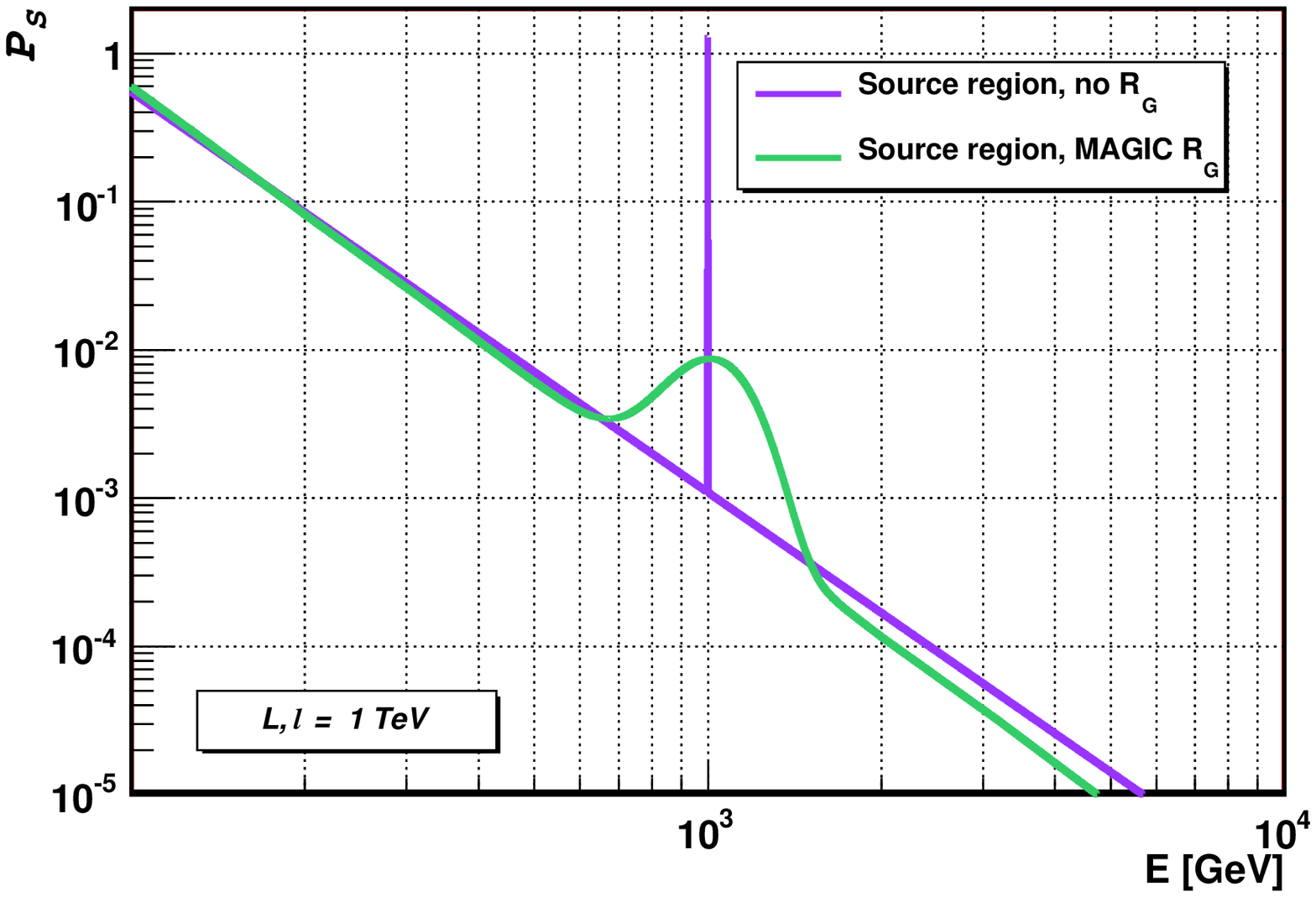}
  \vspace{-10pt}
  \caption{Contribution of the source region to the PDF before 
    (purple) and after the convolution (green) with the response 
    function of the telescope. \textbf{Left}: The spectral slope 
    of a power law-shaped signal is harder after the convolution. 
    \textbf{Right}: A monochromatic line is smoothed and 
    widened due to the finite energy resolution. Shape of the 
    background (\textbf{left} and \textbf{right}) is also affected 
    by the response function.}
  \label{Fig2}
  \vspace{0pt}
\end{figure}

The convolution of spectral and response functions yields the form of the PDF. As seen in figure \ref{Fig2}, the original 
spectral shape is modified by the imperfect instrument, with features like line being smoothed and hardness of the power 
law being altered.

\paragraph{Improvement Factor.}In our tests, we choose the signal intensity ($A_{PL}$ in the case of a $PL$, $A_{L}$ for the 
line-shaped signal) as a free parameter, whose value is to be estimated by maximization of the likelihood functions 
(eq.(\ref{eq1}) and eq.(\ref{eq2})). Performance of the full likelihood with respect to the conventional one, for a given signal 
model $M(\boldsymbol\theta)$, is quantified by means of an \textit{Improvement Factor} ($IF$):
\begin{equation}
IF(M(\boldsymbol\theta)) = \langle CI_{cnvn} / CI_{full} \rangle, 
\label{eq14}
\end{equation}
i.e. the average ratio of the widths of the confidence intervals, $CI_{cnvn}$ and $CI_{full}$, calculated by the corresponding 
methods, assuming a common confidence level.
\begin{figure}[t]\vspace{-20pt}
  \centering
  \includegraphics[width=0.55\textwidth]{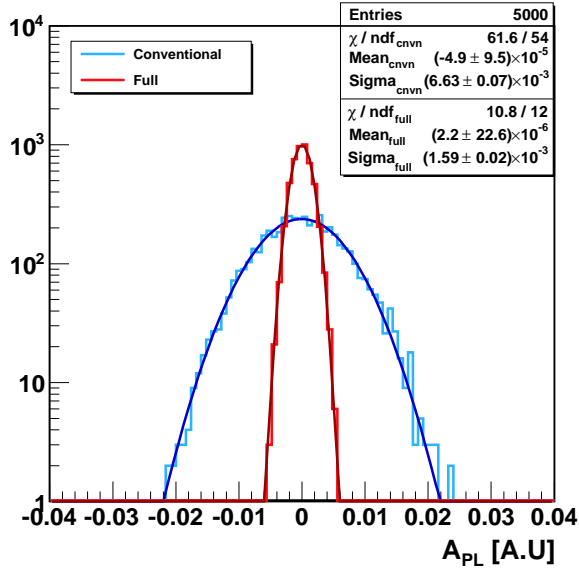}
  \vspace{-15pt}
  \caption{Distribution of the free parameter values 
    estimated by the conventional (blue) and full 
    likelihood method (red), for a $PL$ signal emission 
    model, with $\gamma$ = 1.8. Test conditions are 
    such that the expected parameter value is zero; 
    results are obtained from 5000 fast-simulated 
    experiments.}  
  \vspace{0pt}
  \label{Fig3}
\end{figure} 

The Improvement Factor is, by construction, the \emph{improvement in the sensitivity} of a given search expected by the 
use of the full likelihood over the conventional approach, provided that both methods produce unbiased estimators. We 
have explicitly checked this extreme, for several different models of signal emission, without finding any indications for 
the presence of bias (figure \ref{Fig3}).

In this work, the confidence intervals are two-sided and computed following the ``$ln\mathcal{L}+1/2$'' rule and 
assuming one unconstrained degree of freedom. The maximization of the likelihood functions (eq.(\ref{eq1}) and 
eq.(\ref{eq2})) is performed using the \texttt{TMinuit} class incorporated in the framework of ROOT \cite{3.1.2,3.1.3}. 
We have numerically confirmed that the obtained coverages are the expected ones.

For the characterization of the method, the confidence intervals are calculated with 95\% confidence level, and their ratio 
averaged from 25 fast-simulated experiments. Each simulation consists of $10^5$ events\footnote{The number of events 
chosen for the characterization, for the selected setup, corresponds to $\sim$200 hours of observations. This value 
depends very much on the chosen instrument and applied analysis cuts (in particular on the energy threshold), but does 
not have a significant role in the overall Improvement Factor value (for more details, see table \ref{Table1}, section 
\ref{Sec3.4})}, randomly generated according to the PDF describing the expected background (i.e. the expected value of the 
signal intensity is zero), with $\tau$ = 2, $E_{min}$  = 100 GeV and $E_{max}$ = 10 TeV (unless specified otherwise). 

\subsection{Optimization of the integration range}
\label{Sec3.2}

By definition, the full likelihood takes complete advantage of the signal spectral information, so it makes sense to assume 
that maximal sensitivity with this method is achieved when the whole energy range is considered. For the conventional 
concept, however, this does not have to be the case, especially if some distinctive features are expected in the spectra. 
Here, we study the performance of each method for different energy integration ranges. For a chosen model and a given 
method, the optimal integration range is the one resulting in the best sensitivity.

\begin{figure}[t] \vspace{-20pt} \centering
  \includegraphics[width=1.0\textwidth]{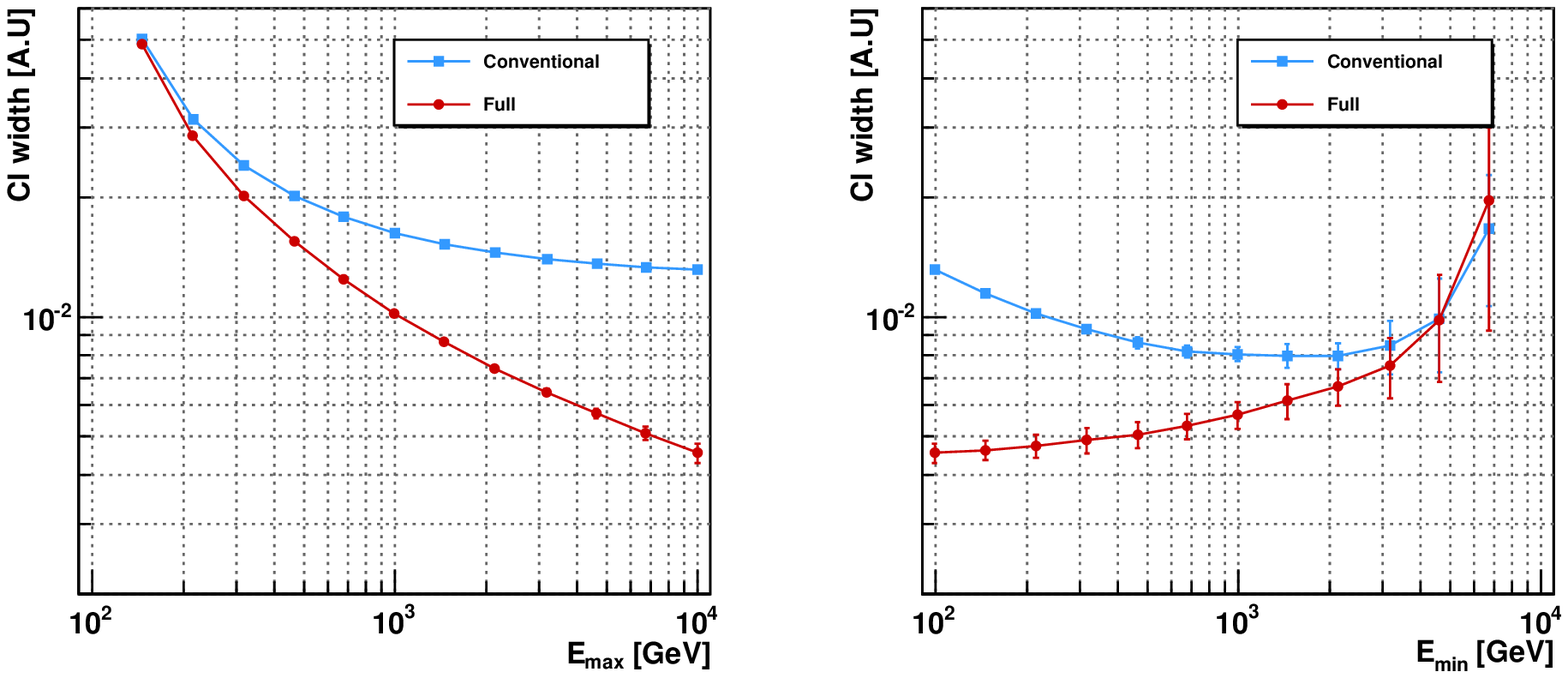}
  \vspace{-25pt}
  \caption{Mean widths of the $CI$s, calculated with 
    the conventional (blue) and full likelihood (red) 
    methods, as a function of the integration range 
    when $E_{min}$ (\textbf{left}) or $E_{max}$ 
    (\textbf{right}) is fixed. The considered signal 
    model is a $PL$ of spectral slope $\gamma$ = 
    1.8. Error bars are the RMS of the $CI$ 
    distributions.}
  \vspace{-5pt}
  \label{Fig4}
\vspace{5pt}
  \includegraphics[width=0.5\textwidth]{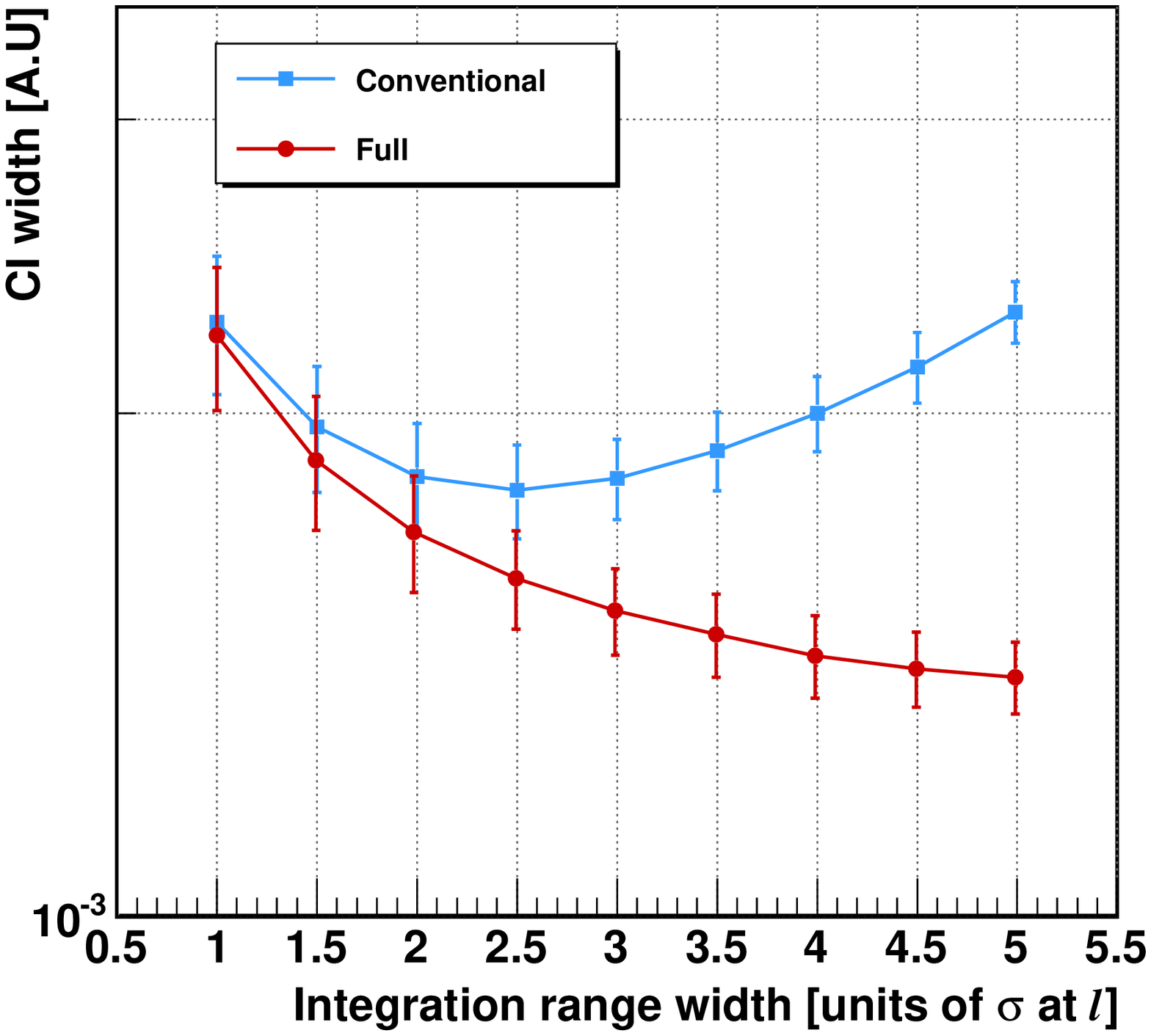}
  \vspace{-15pt}
  \caption{Mean widths of the $CI$s, calculated with 
    the conventional (blue) and full likelihood (red) 
    methods, as a function of the integration range 
    width given in units of $\sigma$ around the line 
    energy $l$ = 1 TeV. Error bars are the RMS of the 
    $CI$ distributions.}
  \vspace{0pt}
  \label{Fig5}
\end{figure}

In the case of the power law-shaped signal, we fix one integration limit while varying the other: figure \ref{Fig4} shows the 
mean values of $CI$s, calculated with each method, for a signal of $\gamma$ = 1.8 spectral slope and the integration 
range of fixed $E_{min}$ (left) or fixed $E_{max}$ (right). As expected, in both cases, the full likelihood is best favoured when 
the entire energy range is considered. As for the conventional approach, the scenario with fixed $E_{max}$ and optimized 
$E_{min}$ yields best sensitivity, and we shall always use such settings in the following tests.

In the case of the spectral line, the sensitivity is optimized by restricting to those events in the vicinity of the peak. Figure 
\ref{Fig5} shows the $CI$ widths of the full likelihood and conventional approaches, as a function of the integration range 
width (expressed in units of $\sigma$), centered at $l$. Again, the full likelihood provides best constraints when the 
complete energy span is integrated, while the conventional approach is most sensitive for a limited range.

The Improvement Factor values given in the following sections are always calculated from the most constraining upper 
limits of both methods, using the whole energy range for the full likelihood and the optimized one for the conventional 
approach.
\subsection{Improvement Factor for different signal models}
\label{Sec3.3}

In this section we compare the sensitivities of the full likelihood and conventional methods for various $PL$ and 
line-shaped signals.

\begin{figure}[t] 
  \vspace{-20pt}
  \centering
  \begin{subfigure}[l]{0.49\textwidth}
    \includegraphics[width=\textwidth]{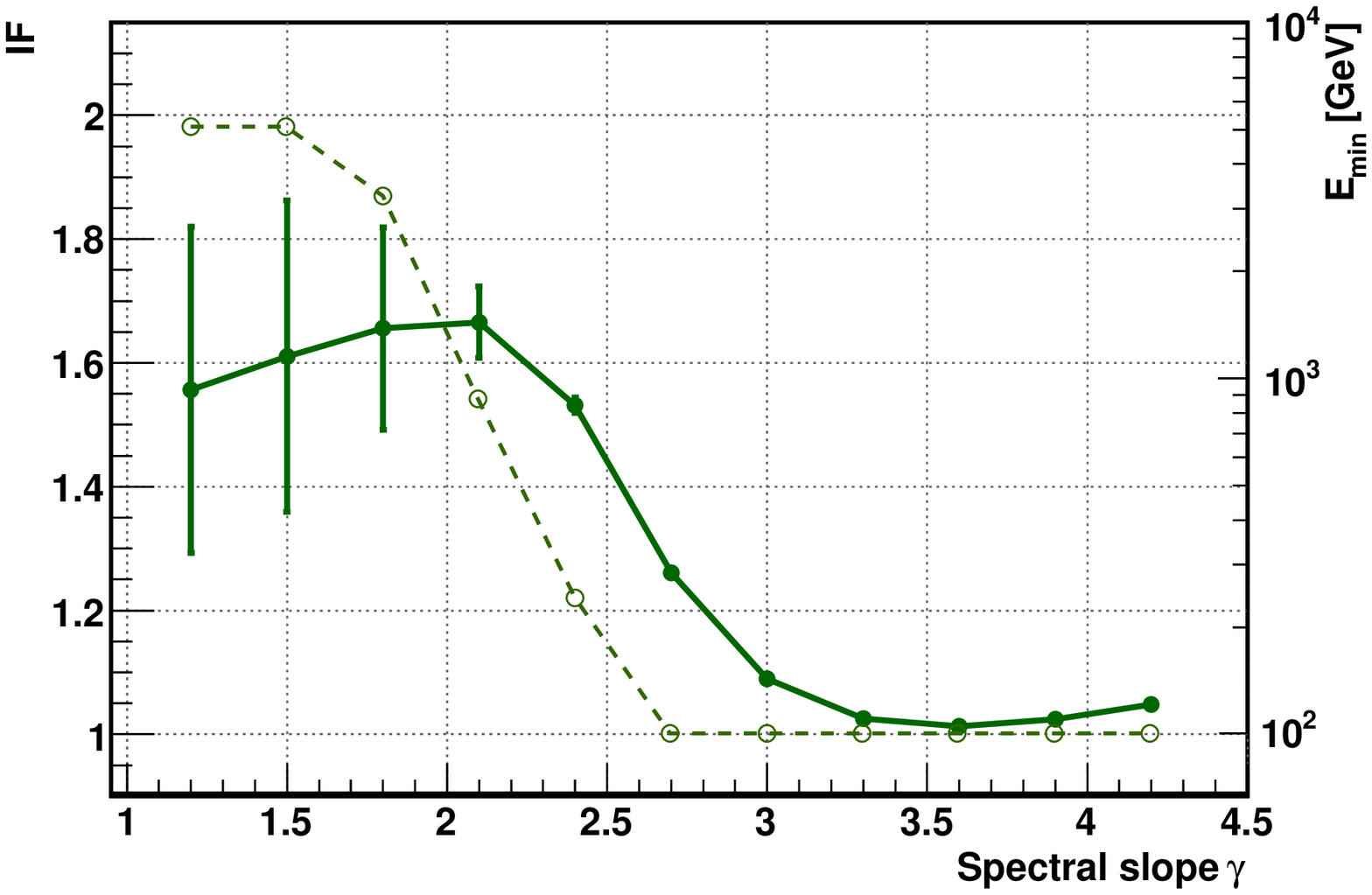}
    \vspace{-25pt}   
    \caption{}
    \label{Fig6a}
  \end{subfigure}
  \begin{subfigure}[r]{0.49\textwidth}
    \includegraphics[width=\textwidth]{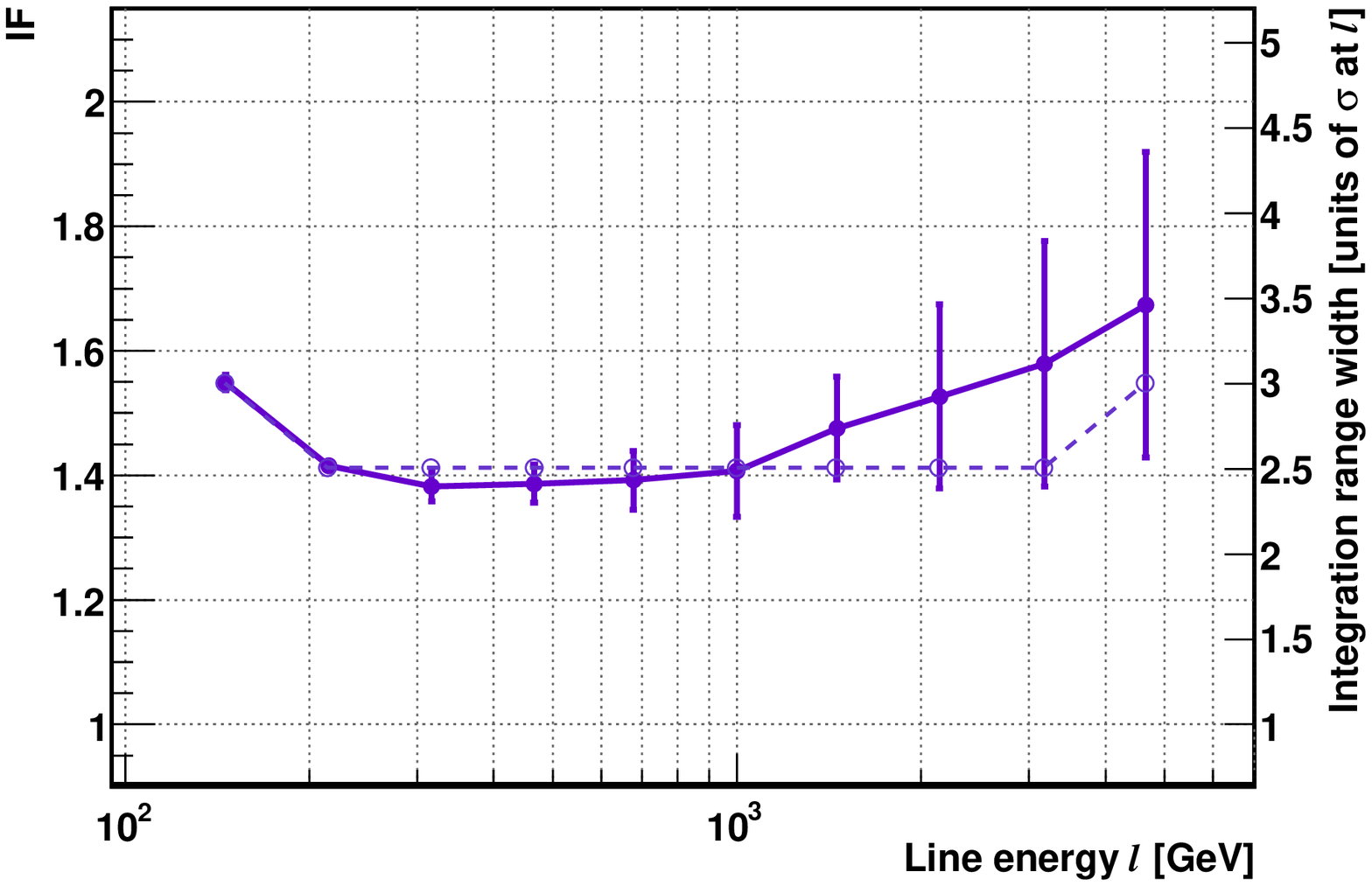}
    \vspace{-25pt}   
    \caption{}
    \label{Fig6b}
  \end{subfigure}
  \vspace{-5pt}
  \caption{Improvement Factor for different $PL$ 
    (\textbf{a}) and $L$ (\textbf{b}) signal models 
    (full line). Also shown are the optimal values of 
    $E_{min}$ and integration range width (conventional 
    approach) for the considered models (dashed line, 
    right-hand axis). Error bars are the RMS of the 
    $IF$ distributions.}
  \vspace{-5pt}
\end{figure}

Figure \ref{Fig6a} shows the Improvement Factor as a function of the spectral slope $\gamma$ for $PL$ models. In this 
example, for the case when $\gamma \approx 3.6$, the shapes of signal and background differential rates are very alike, 
and therefore the improvement one gains from the use of the full likelihood is almost negligible. For harder spectral slopes 
the improvement on sensitivity of the full with respect to that of the conventional likelihood approach can reach up to 
65\%. The dashed line indicates the value of $E_{min}$ for which the conventional method yields the most constraining limit 
for the given model. For expected signal emissions of harder spectral indices, that dominate over the background radiation 
at higher energies, the conventional approach is optimized for the upper end of energy region. For increased $\gamma$, 
differences between signal and background concentrate at lower energies, so integration of the full energy range is 
preferred.
\begin{figure}[t] 
  \vspace{-10pt}
  \centering
  \begin{subfigure}[l]{0.49\textwidth}
    \includegraphics[width=\textwidth]{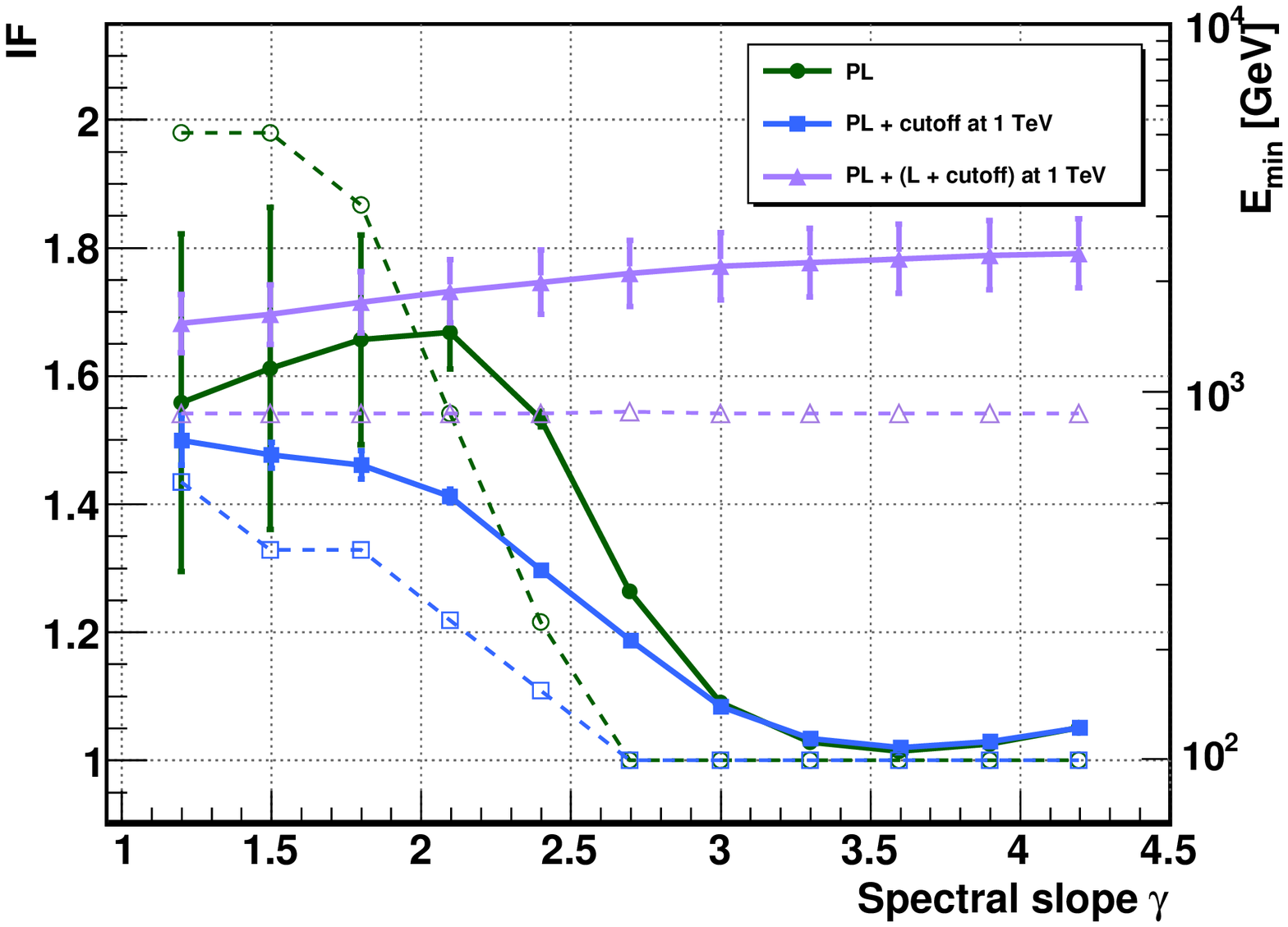}
    \vspace{-25pt}   
    \caption{}
    \label{Fig7a}
  \end{subfigure}
  \begin{subfigure}[r]{0.49\textwidth}
    \includegraphics[width=\textwidth]{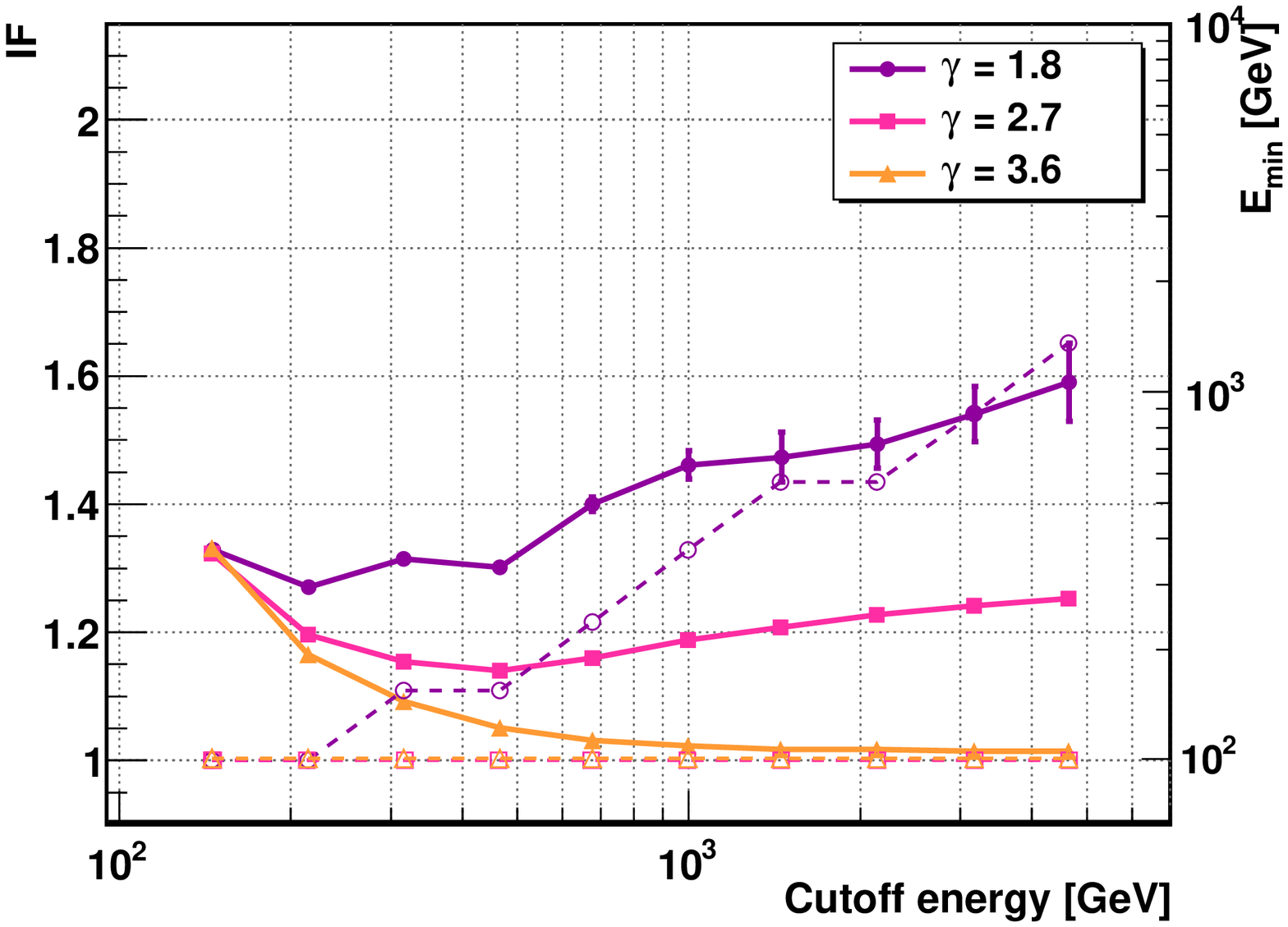} 
    \vspace{-25pt}   
    \caption{}
    \label{Fig7b}
  \end{subfigure}
  \vspace{-5pt}
  \caption{Improvement Factor as a function of spectral 
    slope $\gamma$ (\textbf{a}) and of cutoff energy 
    (\textbf{b}), for different signal models (full lines). 
    Also shown are the optimal values of $E_{min}$ 
    (conventional approach) for the considered models 
    (dashed lines, right-hand axis). Error bars are the RMS 
    of the $IF$ distributions.} 
  \vspace{0pt}
\end{figure}

For $L$ models, depending on the line energy $l$, the Improvement Factor can be between $\sim$40 and 65\% (figure 
\ref{Fig6b}). It is also interesting to note that the width of the optimal integration range (in units of $\sigma$ at $l$) for the 
conventional approach is almost constant for all the models and of order of 2.5 - 3.

We can further elaborate the spectral shape of our signal by including additional features of physical interest. For example, 
the continuous, power law-shaped emission can abruptly cease at a certain energy, resulting in a sharp cutoff in the 
spectral distribution, smoothed by the response function of the detector. Figure \ref{Fig7a} considers the case of $PL$ 
models with different spectral slopes $\gamma$ that all have a cutoff at a fixed energy of 1 TeV. In the presence of a 
cutoff, the Improvement Factor is lower than in the case of uninterrupted $PL$ emission. This is especially noticeable for 
those signal models that dominate at high energies ($\gamma < \alpha$), since their distinction from the background is 
partially erased by the cutoff. For the softer spectra, this effect is not that evident, as for those cases signal is more 
distinguishable from the background at lower energies, i.e. well below the cutoff. 

We also inspect how the Improvement Factor depends on the cutoff energy. Figure \ref{Fig7b} shows that, for hard spectra, 
the higher the cutoff, the greater the improvement. For the soft spectra, Improvement Factor is enhanced by low-energy 
cutoffs to levels comparable to those obtained for spectral lines at similar energies.

Lastly, we study the effect of adding a line to the $PL$-with-cutoff spectral distribution. For such models, we take the 
overall signal intensity as a free parameter, while the individual amplitudes of the $PL$, $A_{PL}$, and $L$, $A_{L}$, are set in 
such a manner that the integrated areas corresponding to those emissions in the PDF are equal. As shown on figure 
\ref{Fig7a}, presence of a line at the same energy as the cutoff ($l$ = 1 TeV) significantly boosts the Improvement Factor 
value, especially for soft spectra. Its contribution is obvious from the optimal $E_{min}$ distribution as well: regardless of 
the value of $\gamma$, the most constraining limits from the conventional method are achieved when $E_{min}$ is just 
below the line, seeing how this feature is the one dominating the Improvement Factor value. 

\subsection{Stability and robustness}
\label{Sec3.4}

While in the previous subsection we varied the signal model, here we analyse the dependence of the Improvement Factor on 
the experimental parameters. Table \ref{Table1} summarizes our results when, for several models of signal emission, we 
take different values of those parameters that are not affiliated with the DM model itself, but rather with the observational 
setup ($\tau$, number of events), energy resolution of the instrument ($\sigma$) and energy range ($E_{max}$). For the 
majority of the considered settings, the observed variations of the Improvement Factor value are no more than 1\% - 2\%, 
with the following exceptions:
\begin{itemize}
\item $\boldsymbol\tau$: the greater the $\tau$, the lower the gain provided by the full likelihood 
method. This is especially noticeable for hard $PL$ and $L$ signal models;
\item $\boldsymbol\sigma$: in the case of the spectral line, the worse the $\sigma$, the greater the 
improvement. It must be clarified, however, that this does not mean that a poor resolution yields more 
constraining upper limits, but that the advantage of the full likelihood approach is more significant;
\item $\boldsymbol{E_{max}}$: for the $PL$ signal models of harder spectra, that dominate over the 
background at higher energies, the increase of energy range means also the greater Improvement 
Factor. On the other hand, for softer $PL$ model, as well as for the considered $L$, change of this 
experimental parameter produces no significant effect. 
\end{itemize}

Additionally, we study the effect the presence of a signal in the data sample may have on the Improvement Factor value. 
Considering various signal intensities, that yield (for full likelihood) significances of up to 5 standard deviations, we find 
that the Improvement Factor increases up to $\sim$10\% for signal model of hard power law spectrum, whereas it has no 
sizable influence for other considered models (table \ref{Table1}).
\begin{table}[t]
  \begin{center}
    \setlength{\extrarowheight}{1pt}
    {\small
      \begin{tabulary}{0.99\textwidth}{  C C C C C }
        \toprule[0.1em]\midrule
        {\multirow{2}{*}{\textbf{Parameter}}} & \textbf{Variation Range} & \multicolumn{3}{c}{\textbf{IF}}\\ 
        & [units of the parameter] & $PL$, $\gamma$ = 1.8 & $PL$, $\gamma$ = 3.6 & $L$, $l$ = 1 TeV \\ \midrule[0.1em]
        $\tau$ & 1 - 5 & 1.91 - 1.47 & 1.02 - 1.01 & 1.63 - 1.26 \\\midrule
        Number of events & $5\times10^4$ - $5\times10^6$ & 1.66 - 1.62 & 1.03 - 1.02 & 1.43 - 1.41 \\\midrule
        $\sigma$ [\% of $\sigma_{MAGIC}$] & 50 - 500 & 1.65 - 1.66 & 1.01 - 1.11 & 1.37 - 2.83\\\midrule
        $E_{max}$ [TeV] & 10 - 50 & 1.65 - 1.82 & 1.01 - 1.02 & 1.40 - 1.41 \\\midrule
        Significance [std. dev] & 0 - 5 & 1.65 - 1.75 & 1.01 - 1.01 & 1.40 - 1.42\\\midrule\bottomrule[0.1em]
      \end{tabulary}
    }
    \vspace{-10pt}
  \end{center}
  \caption{Dependence of the Improvement Factor on 
    different experimental parameters for three different 
    representative signal models.} 
  \label{Table1}
  \vspace{-5pt}
\end{table}

We also test the robustness of the full likelihood by assuming that the response function of the detector is not precisely 
known. For this, we simulate events with one response function, $R_0$, but use a different one, $R_W$, for the likelihood 
maximization. Data are generated so that they contain a gamma-ray signal of intensity that yields a 5$\sigma$ detection 
for $R_0 = R_W$. We study how the significance of the detection by the full likelihood degrades when $R_W \ne R_0$.

First, we consider the effect of using the wrong $A_{eff}$ function, shifted for a fixed value in energy with respect to the real 
one. While for $L$ signal models the sensitivity is not influenced by this discrepancy, for $PL$, especially those of soft 
spectral indices, the sensitivity decreases up to 5\% for a 50 GeV shift.

Next, we consider the case of unknown energy resolution $\sigma$: for a power law-shaped signal, there is no significant 
effect - less than 1\% decline in sensitivity for a factor 2 mistake in the estimate of $\sigma$. On the other hand, in the 
case of a line, a $\sigma$ wrong by the same factor leads to a $\sim$10\% worse sensitivity.

Lastly, we assume different energy bias $\mu$ functions for the simulations and for the likelihood analysis. Our findings 
show that, for $\mu$ values shifted from the actual ones by 1$\sigma$ at the given energy, the sensitivity of the analysis 
decreases $\sim$5\%. If the shift is 2$\sigma$, the decline is $\sim$20\%. This means that, when searching for a line in 
the spectrum, we can take as up to 1$\sigma$ wide steps in our scan without risking a significant sensitivity degradation.

Having in mind that even under these extreme and conservative conditions, the worsening in the sensitivity of the full 
likelihood is still smaller than the improvement one gains from its use (with respect to the conventional approach), we may 
conclude that this method is robust.

As mentioned in section \ref{Sec2}, the background in the source region may be known within some uncertainties. Here we 
estimate the effect this can have on the results of the conventional and full likelihood methods. 

First, we consider energy-dependent differences between the $R_{B}$ functions in the source and background regions, 
parametrized as an extra power law term multiplying the first integral in eq.(\ref{eq6}). Its index is introduced in the 
likelihood functions as a nuisance parameter, with a Gaussian probability distribution of mean 0 and width 0.04 (so that 
maximum deviation of 5\% is achieved at any energy). This results in the sensitivity decrease for both the full likelihood 
and conventional method, but more drastically for the latter one: for the case of the $L$ models as well as the hard 
power law-shaped spectra, results from conventional approach are up to $\sim$50\% less constraining. For the full 
likelihood, the corresponding sensitivity losses are smaller: $\sim$5\%  for $L$ and $\sim$25\% for the $PL$ signal 
models. Soft power law spectra are not affected (less than 1\%), for either of the analysis methods. 

The case of global (normalization) differences is examined by treating $\tau$ as a nuisance parameter, with a Gaussian 
probability distribution of a 5\% width. This leads to significant sensitivity losses for the conventional method: 
$\sim$30\% for $L$ models and $\sim$10\% for hard $PL$ signals. The full likelihood is again far more robust, exhibiting 
almost negligible worsening - less than 2\% for both kinds of signal models. On the other hand, soft $PL$ models result 
problematic for both methods, especially when the spectral shape of the signal is similar to that of the background. The 
conventional approach suffers from up to a factor $\sim$8 worse sensitivity, also for all softer signal models. In the case 
of the full likelihood this is less pronounced (up to a factor $\sim$4 sensitivity worsening), and its power is recovered as 
soon as the shape of the expected signal becomes different from that of the background. This is caused, for both methods, 
by high correlation (up to 0.99) between $\tau$ and signal intensity when the signal and background are of similar spectral 
shapes. For other signal models the correlation is low, due to the energy range optimization applied in the conventional 
approach and the presence of the spectral term in the full likelihood.

\section{Sensitivity of the full likelihood method for Dark Matter searches}
\label{Sec4}

So far we have characterized the performance of the full likelihood in a rather general way, by assuming generic spectral 
shapes. In this section, however, we explore its sensitivity for specific DM models, for the observations with MAGIC and 
CTA\footnote{https://www.cta-observatory.org}. The following cases are considered:
\begin{enumerate}[a)]
  \item benchmark models (BM), as defined by Battaglia et al. (2009) \cite{4.1} and by Bringmann, 
    Doro and Fornasa (2009) \cite{4.2}. We compare our results to those presented in \cite{4.2};
  \item secondary gamma rays from annihilation into SM particles (in particular, we study the 
    $b\bar{b}$, $\tau^{+}\tau^{-}$ and $W^{+}W^{-}$ channels). We also show the sensitivity 
    improvement obtainable through the use of the full likelihood with respect to the recently 
    published VERITAS results of Segue 1 observations \cite{4.3};
  \item annihilation into $\gamma\gamma$. We compare the expected sensitivities to the 
    possible hint of monochromatic line found in Fermi data by Weniger (2012) \cite{4.4} and 
    more recently by Su and Finkbeiner (2012) \cite{4.5}.
\end{enumerate}

To compute the sensitivity of the full likelihood, we use, in eq.(\ref{eq2}), the differential gamma-ray flux from the 
annihilation of DM particles, given as a product of two terms: 
\begin{equation}
\frac{d\Phi_{G}}{dE'} = \frac{d\Phi_{G}^{PP}}{dE'}\times \tilde{J}(\Delta\Omega).
\label{eq15}
\end{equation}
The \emph{particle physics} term, $d\Phi_{G}^{PP}$/$dE'$, describes the characteristics of the chosen DM model:
\begin{equation}
\label{eq16}
\frac{d\Phi_{G}^{PP}}{dE'} = \frac{1}{4\pi} \frac{\langle \sigma v \rangle}{2m_{\chi}^2} \frac{dN_{G}}{dE'},
\end{equation}
where $m_{\chi}$ refers to the DM particle mass, $\langle \sigma v \rangle$ is the thermally averaged cross-section, and 
$dN_{G}$/$dE'$ represents the differential gamma-ray rate per annihilation, summing all possible final states weighted by 
their corresponding branching ratios. 

The effective \emph{astrophysical factor}, $\tilde{J}(\Delta\Omega)$, depends on the distance and morphology of the 
source, and it is defined as the integral along the line of sight ($los$) of the squared DM density $\rho$, integrated over 
the solid angle $\Delta\Omega$ of the signal region: 
\begin{equation}
\label{eq17}
\tilde{J}(\Delta\Omega) = \int_{\Delta\Omega}d\Omega\int_{los}\rho^{2}(r)ds.
\end{equation}

It is reasonable to assume that the $\tilde{J}$ factor, for a given source and assumed DM distribution profile, is the same 
for every IACT of the current generation, since their point-spread functions are very alike, and therefore the signal region 
spans over similar solid angles $\Delta\Omega$. Unlike $\tilde{J}$, $d\Phi_{G}^{PP}$ /$dE'$ does not depend on the 
observed source - its value is completely determined for a given theoretical framework.

In the upcoming subsections, we express the sensitivity of the full likelihood approach as the value of $\langle \sigma v 
\rangle$ (taken as free parameter in the maximization of the likelihood) for which we would obtain a detection with a given 
statistical significance in a given $T_{OBS}$.

\subsection{Benchmark models}
\label{Sec4.1}

Bringmann, Doro and Fornasa (2009) \cite{4.2} made observability predictions (requiring a 5$\sigma$ detection in 50 
hours) for two dwarf Spheroidal galaxies, Draco and Willman 1, for the case of several mSUGRA \cite{4.1.1} BM models, and 
observations with MAGIC and CTA (although, the response functions attributed to each instrument are rather simplified and 
slightly optimistic). In their calculations, they relied on the conventional likelihood approach, and made two studies: one, 
for which $E_{min}$ is the actual energy threshold of the analysis (70 GeV for MAGIC, 30 GeV for CTA), and the other, for 
which $E_{min}$ is optimized for each model based on the sensitivity curves of the instruments. In both cases, $E_{max}$ is 
selected as the DM particle mass $m_{\chi}$.

We have computed the sensitivities of the full and conventional likelihood approaches under the same circumstances 
studied in \cite{4.2}: considering the same DM candidate sources, same BM models of DM emission, and same 
observatories (but with more realistic response functions: the actual one of MAGIC \cite{3.1.1} and one of the latest 
estimates of the response function of the CTA \cite{4.1.2}). The results are shown in table \ref{Table2}, together with the 
basic characteristics of each of the studied BM models. Improvement Factors $IF_{1}$ and $IF_{2}$ represent the gain the full 
likelihood provides over the conventional method, for the two cases of integration ranges considered in \cite{4.2}.

\begin{table}[t]
  \begin{center}
    \setlength{\extrarowheight}{1pt}
    {\small
      \begin{tabulary}{1.00\textwidth}{ C C C C C C C C C }
        \toprule[0.1em]\midrule
        &  & & \multicolumn{3}{c}{\textbf{MAGIC} ($>$70 GeV)\vspace{1.5pt}} & \multicolumn{3}{c}{\hspace{0.4cm}\textbf{CTA} ($>$30 GeV)\vspace{1.5pt}} \\
        \multirow{2}{*}{\textbf{BM}} & \textbf{$m_{\chi}$} & $\sigma v|_{v=0} $ & $\langle \sigma v \rangle_{full}$ & \multirow{2}{*}{$IF_1$} & \multirow{2}{*}{$IF_2$} & $\langle \sigma v \rangle_{full} $ & \multirow{2}{*}{$IF_1$} & \multirow{2}{*}{$IF_2$} \\ 
        & [GeV] & [cm$^3$ s$^{-1}$] & [cm$^3$ s$^{-1}$] & & & [cm$^3$ s$^{-1}$] & & \\ \midrule
        $I'$ & 141 & $3.6\times10^{-27}$& $5.65\times10^{-23}$ & 1.62 & 1.57 & $1.39\times10^{-23}$ & 1.48 & 1.48 \\
        $J'$ & 316 & $3.2\times10^{-28}$ & $1.01\times10^{-23}$ & 3.64 & 1.80 & $1.91\times10^{-24}$ & 5.18 & 1.65 \\
        $K'$ & 565 & $2.6\times10^{-26}$ & $3.91\times10^{-23}$ & 1.23 & 1.23 & $8.39\times10^{-24}$& 1.58 & 1.58 \\
        $BM3$ & 233 & $9.2\times10^{-29}$ & $7.21\times10^{-25}$ & 4.14 & 1.89 & $1.35\times10^{-25}$ & 6.62 & 1.61 \\
        $BM4$ & 1926 & $2.6\times10^{-27}$ & $2.87\times10^{-23}$ & 2.10 & 2.10 & $4.82\times10^{-24}$ & 3.81 & 3.81 \\ \midrule
        \bottomrule[0.1em]
      \end{tabulary}  
    }
    \vspace{-10pt}
  \end{center}
  \caption{Characteristics of the studied BM models 
    (mass $m_{\chi}$ and predicted annihilation cross 
    section today $\sigma v|_{v=0}$), together with the 
    upper limits on the $\langle \sigma v \rangle$ 
    value calculated with full likelihood method 
    ($\langle \sigma v \rangle_{full}$), for Willman 1 
    observations with MAGIC and CTA. We also quote 
    the Improvement Factors obtainable from full 
    likelihood with respect to the conventional 
    approach, computed according to the prescription 
    presented in \cite{4.2}: $IF_1$ is calculated for an 
    integration range (for the conventional method) 
    from energy threshold to $m_{\chi}$, while for 
    $IF_2$ the integration is done from optimized 
    lower limit to $m_{\chi}$.}
  \label{Table2}
\vspace{-5pt}
\end{table}
The lowest Improvement Factors (although of values higher than 25\%) are obtained, for both MAGIC and CTA, for the 
practically featureless, soft spectra of the model $K'$, as well as for the model $I'$ from the bulk region, that has a cutoff 
at low energies. On the other hand, the greatest improvements are achieved in the case of the model $BM4$, characterized 
by the massive DM particle and hard spectrum. Models from coannihilation region, with particularly large internal 
bremsstrahlung contributions, $J'$ and $BM3$, also show significant gain from the use of the full likelihood (above 60\%). 
Despite these high improvements, however, estimated $\langle \sigma v \rangle$ limits are still $\sim$4 and $\sim$3 
orders of magnitude, for MAGIC and CTA respectively, away from the predicted values of these BM models.  

The fact that our results for $\langle \sigma v \rangle_{full}$ are about factor $\sim$2 less constraining than the 
conventional limits presented in \cite{4.2}, can be understood by taking into account that the latter were obtained 
assuming a somewhat idealized situation, with perfectly known background and with flat, optimistic response functions, 
while we consider circumstances of the real experiment and the actual (or latest from the simulations) responses of the 
detectors.

\subsection{Secondary gamma rays from annihilation into SM particles}
\label{Sec4.2}

One of the most recent results from indirect DM searches with IACTs comes from the VERITAS 
Collaboration\footnote{http://veritas.sao.arizona.edu}, from observations of the dwarf spheroidal galaxy Segue 1 \cite{4.3}. 
Although this source is considered to be one of the most DM dominated objects known so far \cite{4.2.1,4.2.2}, nearly 50 
hours of data showed no significant gamma-ray excess. Consequently, upper limits to the $\langle \sigma v \rangle$ 
were derived, for the full energy range and relaying on the conventional likelihood analysis.
\begin{figure} \vspace{-20pt}
  \centering
  \begin{subfigure}[l]{0.49\textwidth}
    \includegraphics[width=\textwidth]{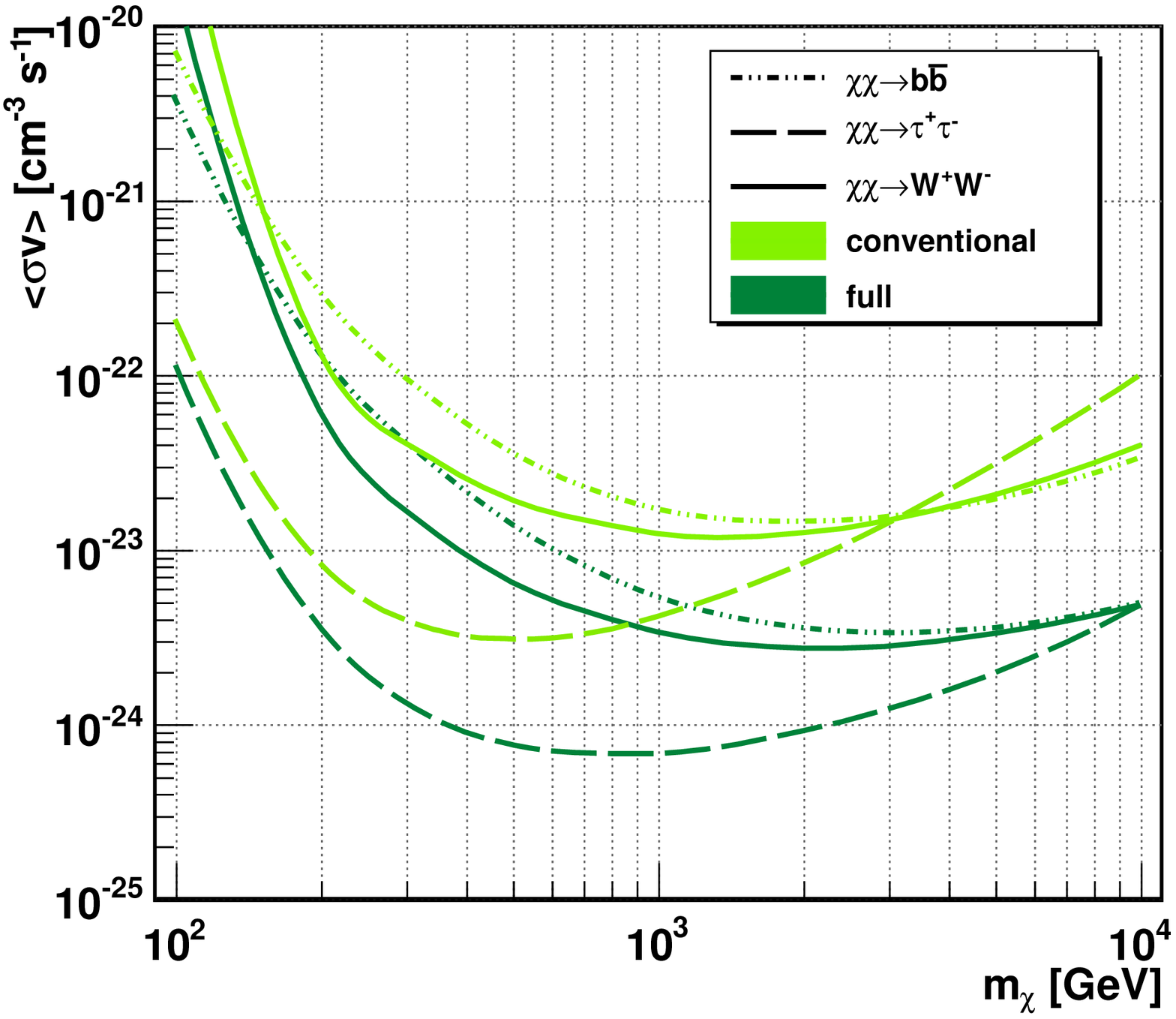}
    \vspace{-25pt}   
    \caption{}
    \label{Fig8a}
  \end{subfigure}
  \begin{subfigure}[r]{0.49\textwidth}
    \includegraphics[width=\textwidth]{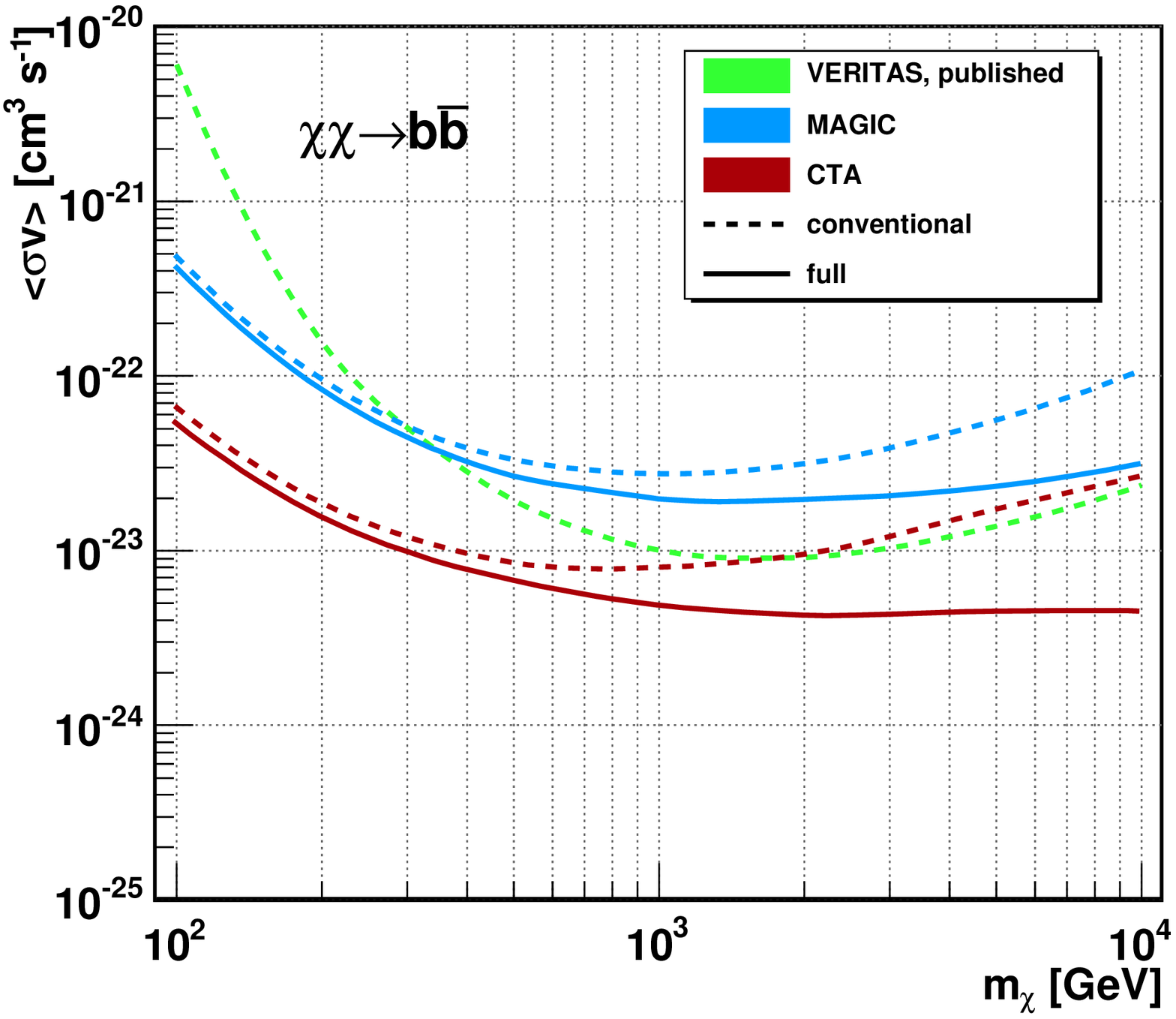}
    \vspace{-25pt}
    \caption{}
    \label{Fig8b}
  \end{subfigure}
  \begin{subfigure}[l]{0.49\textwidth}
    \includegraphics[width=\textwidth]{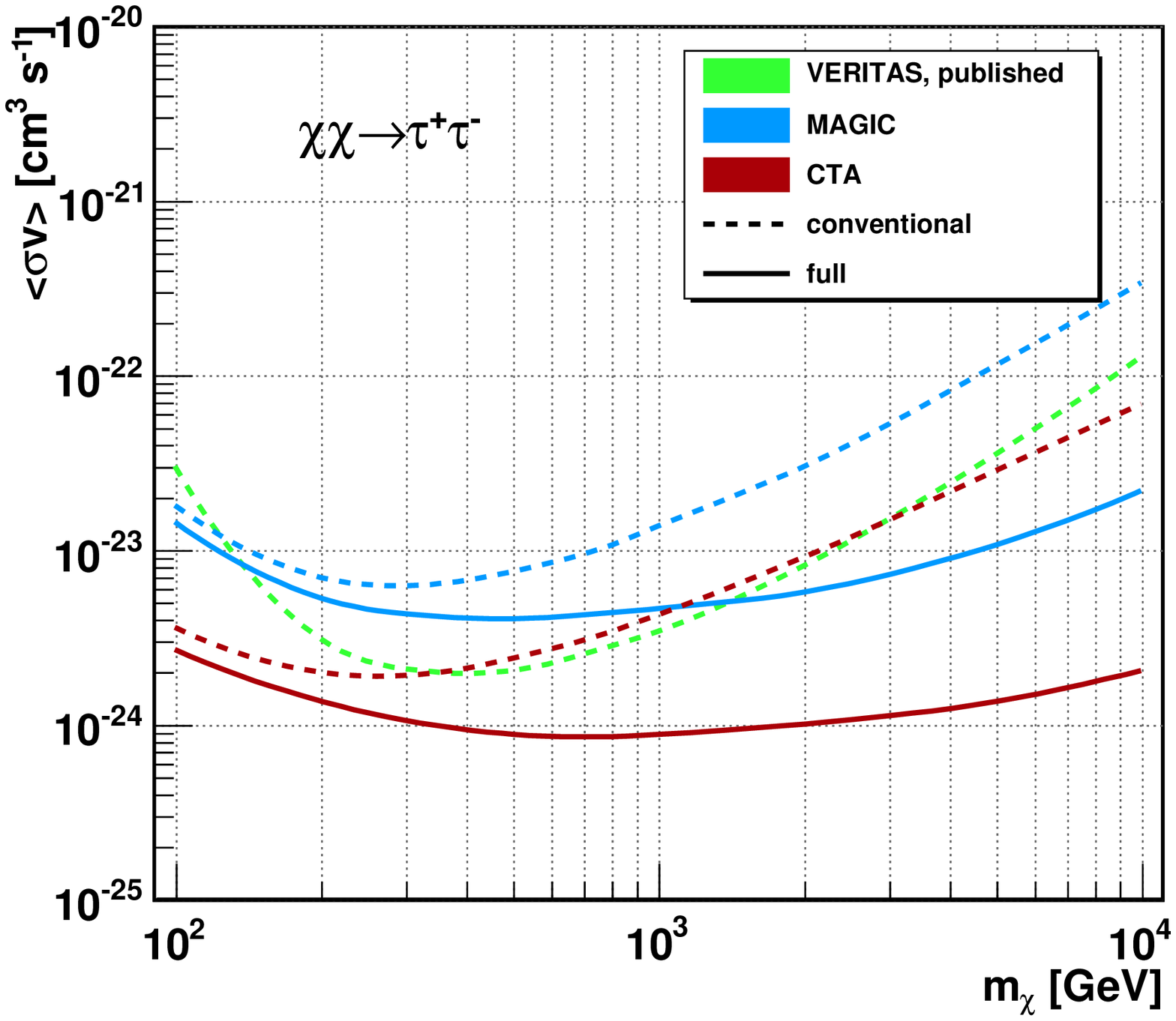}
    \vspace{-25pt}
    \caption{}
    \label{Fig8c}
  \end{subfigure}
  \begin{subfigure}[r]{0.49\textwidth}
    \includegraphics[width=\textwidth]{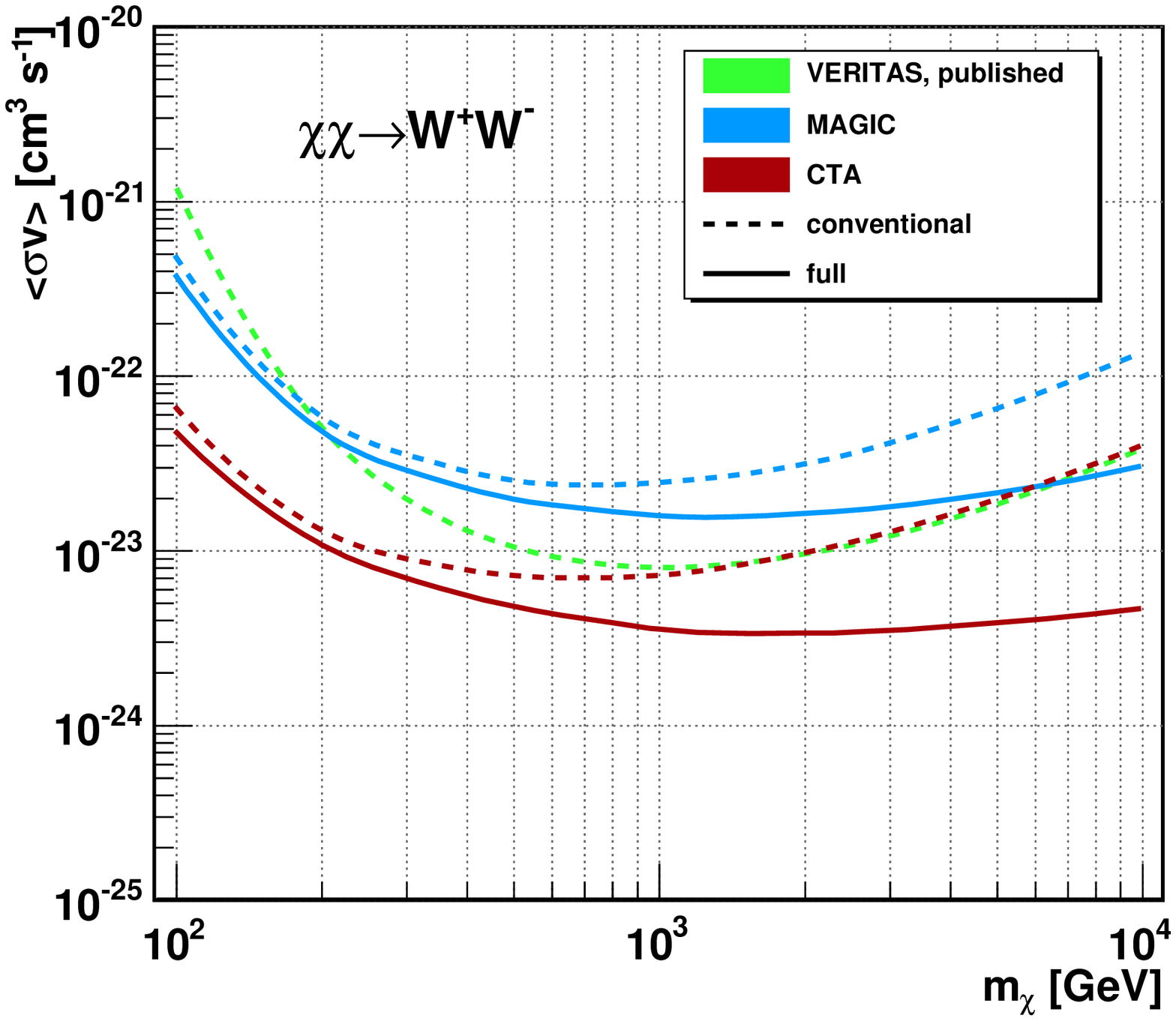}
    \vspace{-25pt}
    \caption{}
    \label{Fig8d}
  \end{subfigure}
  \vspace{-5pt}
  \caption{95\% confidence level upper limits on the 
    $\langle \sigma v \rangle$ as a function of 
    $m_{\chi}$, considering different final state particles, 
    and $\sim$50 h of observations of the Segue 1 
    galaxy. \textbf{(a)} Results for VERITAS, from this 
    work, estimated by the conventional (light green) 
    and from the full likelihood approach (dark green). 
    Exclusion lines for \textbf{(b)} $\chi\chi 
    \rightarrow b\bar{b}$, \textbf{(c)} $\chi\chi 
    \rightarrow \tau^{+}\tau^{-}$ and \textbf{(d)} 
    $\chi\chi \rightarrow W^{+}W^{-}$ channels, for 
    MAGIC (blue) and CTA (red), obtained from both the 
    conventional (dashed line) and full likelihood 
    approach (full line). \textbf{(b-d)} Results from 
    VERITAS, as given in \cite{4.3}, are also plotted for 
    the comparison purposes (green).} 
  \vspace{-0pt}
  \label{Fig8}
\end{figure}

We have computed how these results could be further strengthened if the full likelihood was used instead. For three final 
state channels, $b\bar{b}$, $\tau^{+}\tau^{-}$ and $W^{+}W^{-}$ (100\% branching ratios), we estimate the limits using both 
methods and calculate the Improvement Factor values for the upper limits on the DM particle annihilation cross section 
value $\langle\sigma v\rangle$. Following the prescription from \cite{4.3}, we assume $\tau$ = 1/0.084, $T_{OBS}$ = 47.8 
hours, $\tilde{J}$ = 7.7$\times10^{18}$ GeV$^2$ cm$^{-5}$, and we calculate the upper limits with 95\% confidence level. In 
the lack of the VERITAS response function used for the analysis in \cite{4.3}, we use the $A_{eff}$ function from Wood 
(2010) \cite{4.2.3}, $\sigma$ from \cite{4.2.4} and we assume a conservative $\mu$ = 0. The total background rate is 
taken from \cite{4.3}, approximating the shape of $P_{B}$ dependence on energy by that of MAGIC. For the $dN_{G}/dE'$ of 
the studied channels we use the parametrization from Cembranos et al. (2010) \cite{4.2.5}.

Figure \ref{Fig8a} shows our estimates of $\langle \sigma v \rangle$ limits, calculated by both full and conventional 
likelihood approaches, assuming observations with VERITAS, for considered channels and following the analysis 
prescription as given in \cite{4.3}. The more massive the DM particle, the greater the improvement, especially for the 
$\tau^{+}\tau^{-}$ channel whose spectra gets harder for higher $m_{\chi}$ values. This is partially due to the 
non-optimization of the integration range in the VERITAS analysis. Gain achievable through the use of the full likelihood is 
quite significant: for example, the Improvement Factors for $m_{\chi}$ = 100 GeV, 1 TeV and 10 TeV, are 1.2, 1.6 and 3.4 (for 
the $b\bar{b}$ channel), 1.2, 2.9 and 10.1 (for the $\tau^{+}\tau^{-}$) and 1.3, 1.5 and 4.5 (for the $W^{+}W^{-}$ channel), 
respectively. Our estimates do not rely on the actual response function used in \cite{4.3}, and our limits from the 
conventional likelihood are slightly less constraining than those reported by VERITAS. Nevertheless, we confirm that the 
consistent Improvement Factor results ($<5\%$ difference) are obtained when different $A_{eff}$ function is used 
(McCutcheon (2012) \cite{4.2.6}). From this, we infer the validity of the obtained Improvement Factor values also for the 
response function actually applied in the analysis from \cite{4.3}.

Additionally, we study the sensitivities of MAGIC and CTA for the gamma-ray spectra from $b\bar{b}$, $\tau^{+}\tau^{-}$ 
and $W^{+}W^{-}$ channels, assuming the same observational and analysis conditions as in the case of VERITAS, but using 
the actual/simulated response functions of these instruments. Exclusion lines, calculated by means of both full likelihood 
and conventional methods, are shown on figure \ref{Fig8b}-\ref{Fig8d}. As expected, the CTA results are always better than 
those of MAGIC: at lower energies, by a factor $\sim$5, and by more than one order of magnitude at high energies. Again, 
the constraints from the full likelihood approach are more significantly improved with respect to the conventional ones for 
more massive DM particles, and the lowest $\langle \sigma v \rangle$ limits are achieved for the $\tau^{+}\tau^{-}$ channel 
(hardest spectrum). For MAGIC, Improvement Factors for DM particle masses of $m_{\chi}$  = 100 GeV, 1 TeV and 10 TeV are 
rather relevant: 1.2, 1.4 and 3.4 ($b\bar{b}$), 1.3, 2.9 and 15.8 ($\tau^{+}\tau^{-}$) and 1.3, 1.5 and 4.5 ($W^{+}W^{-}$). In the 
case of CTA, these values are even more significant: 1.3, 1.7 and 6.0 ($b\bar{b}$ channel), 1.3, 4.8 and 33.1 
($\tau^{+}\tau^{-}$ channel) and 1.4, 2.0 and 8.5 ($W^{+}W^{-}$ channel), respectively. 

According to the results shown on figure \ref{Fig8b}-\ref{Fig8d}, the sensitivity gain of the CTA with respect to VERITAS 
would be marginal, or even nonexistent, for certain annihilation channels and mass ranges. We have traced this 
inconsistency down to a probable overestimation of the VERITAS performance assumed in \cite{4.3}. For that, we have used 
the response functions assumed for VERITAS and those of MAGIC and CTA to compute the integral sensitivity (5$\sigma$ 
significance in 50 hours of observations) for a Crab-like spectrum\footnote{$dN/dE = 5.8\times10^{-13}(E/300 
\textrm{GeV})^{-2.32-0.13\log_{10}(E/300 \textrm{GeV})}$ GeV$^{-1}$ cm$^{-2}$ s$^{-1}$ \cite{3.1.1}} at the analysis 
threshold\footnote{Defined as the peak of gamma-ray rate true energy distribution}, for the different instruments. The 
results obtained for MAGIC (1.3$\%$ of Crab flux above 110 GeV) and CTA (0.30$\%$ of Crab flux above 75 GeV) are 
consistent with those published by the respective Collaborations (\cite{3.1.1},\cite{4.1.2}). On the other hand, our results 
for VERITAS, estimated assuming the $A_{eff}$ from \cite {4.2.3}, imply a sensitivity of 0.32$\%$ of Crab flux above 165 
GeV, more than a factor 2 better than the one reported at \cite{4.2.7}. And given how the DM constraints reported in 
\cite{4.3} are stronger than those computed in this work, we can expect this discrepancy to be even larger for the $A_{eff}$ 
actually used in the analysis by the VERITAS Collaboration.

\subsection{Annihilation into $\gamma\gamma$}
\label{Sec4.3}

Although theoretically highly suppressed, direct annihilation of the DM particles into gamma rays would result in a presence 
of a sharp line in the energy spectrum, whose detection would represent an unambiguous confirmation of the DM existence. 

\begin{figure} 
  \vspace{-20pt}
  \begin{center}
    \includegraphics[width=0.89\textwidth]{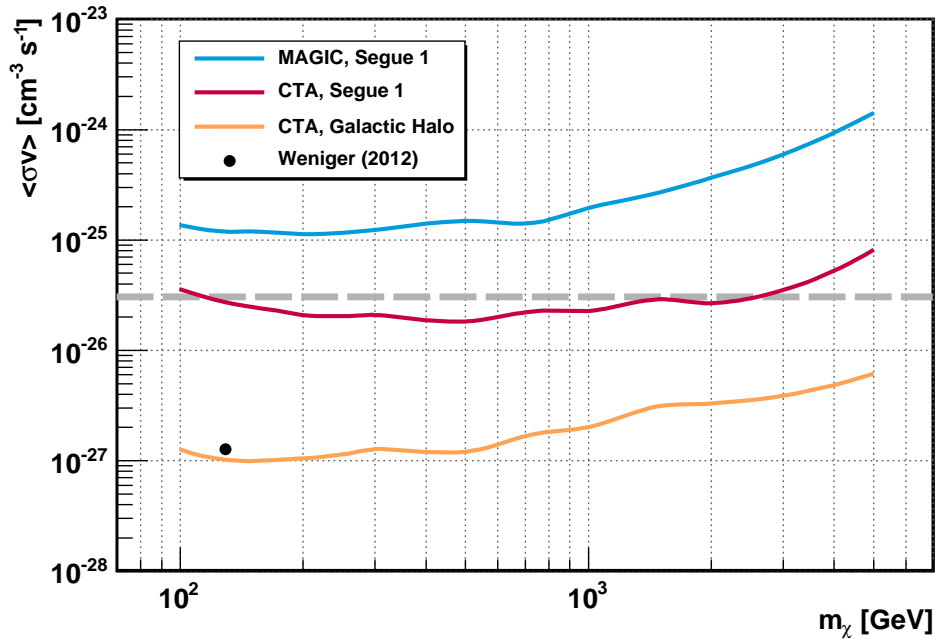}
  \end{center} 
  \vspace{-20pt}
  \caption{MAGIC (blue) and CTA (red) sensitivities 
    on the $\langle \sigma v \rangle$ values as a 
    function of $m_{\chi}$, for 50 hours of Segue 1 
    observations, assuming a 100\% branching ratio 
    into $\gamma\gamma$. Also plotted are the 
    sensitivity predictions for the CTA observations 
    of the Galactic Halo (orange), as well as the 
    $\langle \sigma v \rangle$ value corresponding 
    to DM signal hint, at $m_{\chi}$ = 129 GeV, 
    claimed in \cite{4.4}.}
  \vspace{-0pt}
  \label{Fig9}
\end{figure}

We have made estimates, using the full likelihood approach, of the sensitivity MAGIC and CTA observatories have to spectral 
lines, for the DM particle mass in the energy range between 100 GeV and 5 TeV. We require a 5$\sigma$ detection in 50 
hours of observations of Segue 1, with Einasto\cite{4.3.1} DM profile ($\tilde{J} = 1.7\times10^{19}$ GeV$^2$ cm$^{-5}$ 
\cite{4.3.2}). For the size of the background region we take $\tau$ = 12. Results are shown on figure \ref{Fig9}. MAGIC and 
CTA exhibit greatest sensitivity for $m_{\chi}$ around 200 and 500 GeV, respectively, with CTA being a factor between 
$\sim$5 and $\sim$10 better than MAGIC. Furthermore, for the greater part of the considered $m_{\chi}$ space, the CTA even 
slightly probes the $\langle \sigma v \rangle$ region below the weak-scale cross-section value, $\sim3\times10^{-26}$ 
cm$^{3}$ s$^{-1}$.  

Recent work from Weniger (2012) \cite{4.4} claims a hint of a gamma-ray line at energy of $\sim$129 GeV, from 43 months 
of Galactic Halo public Fermi data and search restricted to the 20 - 300 GeV range. The inferred value of $\langle \sigma v 
\rangle$ is 1.27$\times10^{-27}$ cm$^3$ s$^{-1}$ (assuming Einasto profile). As seen from figure \ref{Fig9}, neither MAGIC 
nor CTA are close to reaching that sensitivity using Segue 1 observations.

On the other hand, observations of a source of much higher $\tilde{J}$ should result in better constraints. For example, 50 
hours of CTA observations of the Galactic Halo (assuming NFW density profile, $\tilde{J} = 3.3\times10^{21}$  GeV$^2$ 
cm$^{-5}$  and $\tau$ = 2) would yield order of 30 times better sensitivity than in the case of Segue 1 (figure \ref{Fig9}). 
Furthermore, 50 hours of data would be sufficient for the  CTA to test Weniger's claim. It must be noted, however, that this 
computation does not take into account the systematic uncertainties, which might be relevant for this search (given how 
the gamma-ray rate would be $\sim$2\% that of the background).

\section{Discussion}
\label{Sec5}

We have presented an analysis approach for IACTs that uses the full likelihood method, constructed to take the maximal 
advantage of the unique spectral features of DM origin. Almost solely through the inclusion of the a priori knowledge on 
the expected gamma-ray spectrum in the likelihood, this method accedes better sensitivity of the analysis, with 
Improvement Factors reaching values up to order of 10 (depending on the signal model) with respect to the recent IACT 
results. In addition, as shown in section \ref{Sec3.4}, these improvements are rather insensitive to other analysis 
characteristics, like the background estimation or signal-to-background discrimination. As a result, the full likelihood can 
be combined to any other analysis developments aimed at further sensitivity enhancements. 

In this work, we have focused on the indirect searches for DM annihilation signals with IACTs. This is reflected in the 
specific form of the likelihood function (eq.(\ref{eq2})), determined by the fact that IACT observations are pointed, cover a 
relatively narrow field of view, and are dominated by background events. Although, to our knowledge, never used for IACTs, 
this concept is a well known analysis method, successfully applied in other fields, including DM searches with different 
techniques and instruments. For instance, a similar approach is employed in the direct detection experiments, like 
XENON100 \cite{5.1}, and even more extensively, in the indirect searches for DM signals in gamma rays by the 
Fermi-LAT\footnote{http://fermi.gsfc.nasa.gov} (see, e.g., \cite{5.2,5.3,5.4}).

The proposed method is sufficiently general to be used in studies of other physics cases studied by the IACTs, the only 
condition being that a prediction about the expected spectral distribution can be made. A trivial example is the search for 
DM decay signals, for which we only need to substitute the $\rho^2$ term by $\rho$ in eq.(\ref{eq17}) and $\langle \sigma 
v \rangle$/$2 m_{\chi}^2$ by $1/\tau_{\chi}m_{\chi}$ in eq.(\ref{eq16}) (with $\tau_{\chi}$ referring  to the DM particle decay 
lifetime). Another example where the full likelihood can be successfully applied is in the search of the AGN spectra for 
signatures induced by the oscillations of gammas into axion-like particles in the presence of intergalactic magnetic fields 
\cite{5.5}. This case, however, would require the a priori assumptions on the AGN emission and effects of gamma rays 
interacting with the extragalactic background light. 

Very important characteristic of the full likelihood method (and any likelihood function-based analysis) is that it allows a 
rather straightforward combination of the results obtained by different instruments and from different targets. For a given 
DM model $M(\boldsymbol\theta)$, and  $N_\textrm{inst}$ different instruments (or measurements), a global likelihood 
function can be simply written as:
\begin{equation}
{\cal L}_T(M(\boldsymbol\theta)) = \prod_{i=1}^{N_\textrm{inst}} {\cal L}_i (M(\boldsymbol\theta)). 
\label{eq18}
\end{equation}
This approach eliminates the complexity required for a common treatment of data and response functions of different 
telescopes or analyses, required by, e.g. the data stacking method (see, e.g. \cite{5.6}). Within the likelihood scheme, the 
details of each experiment do not need to be combined or averaged. The only necessary information is the value of the 
likelihood, expressed as a function of the free parameter (e.g. $\langle \sigma v \rangle$) of a given model for different 
instruments. Since DM signals are universal and do not depend on the observed target, the results from different sources 
can also be combined through the overall likelihood function (as done by Fermi \cite{5.7}), providing therefore a more 
sensitive DM search. For example, combined results (of similar sensitivities to $\langle \sigma v \rangle$) from three 
different observatories (e.g. MAGIC, VERITAS and HESS\footnote{http://www.mpi-hd.mpg.de/hfm/HESS}) would benefit from 
an extra improvement in the sensitivity by a factor of $\sim70\%$. This approach would offer the best chances of 
discovering DM in indirect VHE gamma-ray searches or of setting the most stringent limits attainable by this kind of 
observations, placing therefore a new landmark in the field.  

\paragraph {Acknowledgements}
Authors wish to give thanks to Michele Doro,  Julian Sitarek, Victor Stamatescu and especially 
to Abelardo Moralejo, for providing some needed data and for very useful discussions. This 
work has been supported by Spanish MICINN through the programs FPA (grants FPA2009-07474 
and FPA2010-22056-C06-01) and Consolider-Ingenio-2010 (grant MultiDark CSD2009-00064).



\begin{thebibliography}{00}
\bibitem{1.1} G. Bertone, D.Hooper and J. Silk, \emph{Particle Dark Matter: Evidence, candidates and constraints}, \emph{Phys. Rept.} \textbf{405} (2005) 279-390 [\verb+arXiv:hep-ph/0404175+]
\bibitem{1.2} M. Taoso, G. Bertone and A. Masiero, \emph{Dark Matter Candidates: A ten-point test}, \emph{JCAP} \textbf{03} (2008) 022 [\verb+arXiv:0711.4996+]
\bibitem{1.3} J. Wess and B. Zumino, \emph{Supergauge transformations in four dimensions}, \emph{Nucl. Phys.} \textbf{B70} (1974) 39
\bibitem{1.4} J. L. Feng, \emph{Dark Matter Candidates from Particle Physics and Methods of Detection}, \emph{Ann.Rev. Astron. Astrophys.} \textbf{48} (2010) 495 [\verb+arXiv:1003.0904+]
\bibitem{1.5} A. Birkedal, K. T. Matchev, M. Perelstein and A. Spray, \emph{Robust gamma ray signature of WIMP dark matter}, (2005) [\verb+arXiv:hep-ph/0507194+]
\bibitem{1.6} L. Bergstr\"om and P. Ullio, \emph{Full one-loop calculation of neutralino annihilation into two photons}, \emph{Nucl. Phys.} \textbf{B504} (1997) 27-44 [\verb+arXiv:hep-ph/9706232+]
\bibitem{1.7} T. Bringmann, L. Bergstr\"om and J. Edsj\"o, \emph{New Gamma-ray Contributions to Supersymmetric Dark Matter Annihilation}, \emph{JHEP} \textbf{01} (2008) 049 [\verb+arXiv:0710.3169+]
\bibitem{1.8} T. Bringmann et al., \emph{Fermi LAT Search for Internal Bremsstrahlung Signatures from Dark Matter Annihilation}, \emph{JCAP} \textbf{07} (2012) 054 [\verb+arXiv:1203.1312+]
\bibitem{2.1} W. A. Rolke, A. M. L\'opez and J. Conrad, \emph{Limits and confidence intervals in the presence of nuisance parameters}, \emph{Nucl. Instrum. Methods Phys. Res. A}, Vol. \textbf{551} (2005) 493-503 [\verb+arXiv:physics/0403059+]
\bibitem{2.2} A. Pinzke, C. Pfrommer and L. Bergstrom, \emph{Prospects of detecting gamma-ray emission from galaxy clusters: cosmic rays and dark matter annihilations}, \emph{Phys. Rev. D} \textbf{84} (2012) 123509 [\verb+arXiv:1105.3240+]
\bibitem{2.3} J. F. Navarro, C. S. Frenk and S. D. White, \emph{The Structure of Cold Dark Matter Halos}, \emph{Astrophys. J.} \textbf{462} (1996) 563 [\verb+arXiv:astro-ph/9508025+]
\bibitem{2.4} B. Moore et al., \emph{Cold collapse and the core catastrophe}, \emph{Mon. Not. Roy. Astron. Soc.} \textbf{310} (1999) [\verb+arXiv:astro-ph/9903164+]
\bibitem{2.5} L. Bergst\"om, P. Ullio and J. H. Buckley, \emph{Observability of Gamma Rays from Dark Matter Neutralino Annihilations in the Milky Way Halo}, \emph{Astropart. Phys.} \textbf{9} (1998) 137 [\verb+arXiv:astro-ph/9712318+]
\bibitem{2.6} H. Dickinson and J. Conrad, \emph{Combined Source Analyses for Atmospheric Cherenkov Telescopes}, (2012) [\verb+arXiv:1203.5643+]
\bibitem{3.1.1} MAGIC Collaboration, J. Aleksi\'c et al., \emph{Performance of the MAGIC stereo system obtained with the Crab Nebula data}, \emph{Astropart. Phys.} \textbf {35} (2012) 435-448 [\verb+arXiv:1108.1477+]
\bibitem{3.1.2} F. James, \emph{MINUIT. Function Minimization and Error Analysis, Reference Manual Version 94.1}, \emph{CERN Program Library Long Writeup D506}, CERN Geneva (1994)
\bibitem{3.1.3} \verb+http://root.cern.ch/root/html/TMinuit.html+
\bibitem{4.1} M. Battaglia et al., \emph{Updated post-WMAP benchmarks for supersymmetry}, \emph{Eur. Phys. J.} \textbf{C33} (2004) 273–296 [\verb+arXiv:hep-ph/0306219+]
\bibitem{4.2} T. Bringmann, M. Doro, and M. Fornasa, \emph{Dark Matter Signals from Draco and Willman 1: Prospects for MAGIC II and CTA}, \emph{JCAP} \textbf{01} (2009) 16 [\verb+arXiv:0809.2269+]
\bibitem{4.3} VERITAS Collaboration, E. Aliu et al., \emph{VERITAS Deep Observations of the Dwarf Spheroidal Galaxy Segue 1}, \emph{Phys. Rev. D} \textbf{85} (2012) 062001 [\verb+arxiv:1202.2144+]
\bibitem{4.4} C. Weniger, \emph{A Tentative Gamma-Ray Line from Dark Matter Annihilation at the Fermi Large Area Telescope}, \emph{JCAP} \textbf{08} (2012) 007 [\verb+arXiv:1204.2797+]
\bibitem{4.5} M. Su and D. P. Finkbeiner, \emph{Strong Evidence for Gamma-ray Line Emission from the Inner Galaxy}, (2012) [\verb+arXiv:1206.1616+]
\bibitem{4.1.1} H. P. Nilles, \emph{Supersymmetry, Supergravity and Particle Physics}, \emph{Phys. Rep.} \textbf{110} (1984) 1-162
\bibitem{4.1.2} The CTA Consortium, M. Actis et al., \emph{Design concepts for the Cherenkov Telescope Array CTA: an advanced facility for ground-based high-energy gamma-ray astronomy}, \emph{Exp. Astron.} \textbf{32} (2011) 193-316 [\verb+arXiv:1008.3703+]
\bibitem{4.2.1} M. Geha et al., \emph{The Least-Luminous Galaxy: Spectroscopy of the Milky Way Satellite Segue 1}, \emph{Astrophys. J.} \textbf{692} (2009) 1464-1475 [\verb+arXiv:0809.2781+]
\bibitem{4.2.2} R. Essig et al., \emph{Indirect Dark Matter Detection Limits from the Ulta-Faint milky Way Satellite Segue 1}, \emph{Phys. Rev. D} \textbf{82} (2010) 123503 [\verb+arXiv:1007.4199+]
\bibitem{4.2.3} M. Wood, 2010. \emph{An Indirect Search for Dark Matter with VERITAS}. Ph.D. University of California
\bibitem{4.2.4} J. Holder, 2011, \emph{VERITAS: Status and Highlights}, 32nd ICRC, Bejing, China
\bibitem{4.2.5} J. A. R. Cembranos et al., \emph{Photon spectra from WIMP annihilation}, \emph{Phys. Rev. D} \textbf{83} (2011) 083507 [\verb+arXiv:1009.4936+]
\bibitem{4.2.6} M. McCutcheon, 2012. \emph{Search for VHE gamma-ray emission from the globular cluster M13 with VERITAS}. Ph.D. McGill University
\bibitem{4.2.7} \verb+http://veritas.sao.arizona.edu/about-veritas-mainmenu-81/veritas-specifications+ \\\verb+-mainmenu-111+
\bibitem{4.3.1} J. F. Navarro et al., \emph{The diversity and similarity of simulated cold dark matter haloes}, \emph{Mon. Not. Roy. Astron. Soc} \textbf{402} (2010) 21-34 [\verb+arXiv:0810.1522+]
\bibitem{4.3.2} MAGIC Collaboration, J. Aleksi\'c et al., \emph{Searches for Dark Matter Annihilation Signatures in the Segue 1 satellite galaxy with the MAGIC-I Telescope}, \emph{JCAP} \textbf{06} (2011) 035 [\verb+arXiv:1103.0477+]
\bibitem{5.1} XENON100 Collaboration, E. Aprile et al., \emph{Likelihood Approach to the First Dark Matter Results from XENON100}, \emph{Phys. Rev. D} \textbf{84} (2011) 052003 [\verb+arXiv:1103.0303+]
\bibitem{5.2} Fermi-LAT Collaboration, M. Ackermann et al., \emph{Constraints on Dark Matter Annihilation in Clusters of Galaxies with the Fermi Large Area Telescope}, \emph{JCAP} \textbf{05} (2010) 025 [\verb+arXiv:1002.2239+]
\bibitem{5.3} Fermi-LAT Collaboration, A. A. Abdo et al., \emph{Observations of Milky Way Dwarf Spheroidal Galaxies with the Fermi-Large Area Telescope Detector and Constrains on Dark Matter Halo Models}, \emph{Astrophys. J.} \textbf{712} (2010) 147-158 [\verb+arXiv:1001.4531+]
\bibitem{5.4} Fermi-LAT Collaboration, A. A. Abdo et al., \emph{Fermi Large Area Telescope Search for Photon Lines from 30 to 200 GeV and Dark Matter Implications}, \emph{Phys. Rev. Lett.} \textbf{104} (2010) 091302 [\verb+arXiv:1001.4836+]
\bibitem{5.5} M. A. Sanchez-Conde et al., \emph{Hints of the existence of Axion-Like-Particles from the gamma-ray spectra of cosmological sources}, \emph{Phys. Rev. D} \textbf{79} (2009) 123511 [\verb+arXiv:0905.3270+]
\bibitem{5.6} MAGIC Collaboration, J. Aleksi\'c et al., \emph{Gamma-ray excess from a stacked sample of high- and intermediate-frequency peaked blazars observed with the MAGIC telescope}, \emph{Astrophys. J.} \textbf{729} (2011) 115 [\verb+arXiv:1002.2951+]
\bibitem{5.7} Fermi-LAT Collaboration, M. Ackermann et al., \emph{Constraining Dark Matter Models from a Combined Analysis of Milky Way Satellites with Fermi Large Area Telescope}, \emph{Phys. Rev. Lett} \textbf{107} (2011) 241302 [\verb+arXiv:1108.3546+]
\end{thebibliography}
\end{document}